\RequirePackage{fix-cm}
\documentclass[natbib]{svjour3}                     
\smartqed  
\usepackage{graphicx,subfigure}
 \usepackage{mathptmx}      
\usepackage{bm}
\usepackage{amssymb}
\usepackage{amsmath}
 \journalname{Space Science Reviews} 
\newcommand{\aap}{{Astron. Astrophys.}}
\newcommand{\apj}{{Astrophys. J.}}
\newcommand{\apjl}{{Astrophys. J. Lett.}}
\newcommand{\apjs}{{Astrophys. J. Suppl. Ser.}}

\newcommand{\nat}{{Nature}}
\newcommand{\mnras}{{MNRAS}}
\newcommand{\pasj}{{PASJ}}
\newcommand{\apss}{{Astrophysics and Space Science}}
\newcommand{\memsai}{{Memorie della Societa Astronomica Italiana}}
\newcommand{\aapr}{{The Astronomy and Astrophysics Review}}
\newcommand{\araa}{ARA\&A}
\newcommand{\prd}{Phys.~Rev.~D}
\def\pdot {\dot P}
\def\nudot {\dot \nu}

\def\msun{~M_{\odot}}

\def \curlB {\vec{\nabla}\times (e^\nu \vec{B})}
\def\ltsima{$\; \buildrel < \over \sim \;$}
\def\lsim{\lower.5ex\hbox{\ltsima}}
\def\gtsima{$\; \buildrel > \over \sim \;$}
\def\gsim{\lower.5ex\hbox{\gtsima}}
\def\smc   {CXOU~J0100$-$7211}
\def\uu      {4U~0142$+$61}
\def\zeroq {SGR~0418$+$5729}
\def\zeroc {SGR~0501$+$4516}
\def\lmc    {SGR~0526$-$66}
\def\oo     {1E~1048.1$-$5937}
\def\qui    {1E 1547.0$-$5408}
\def\psr   {PSR J1622$-$4950}

\def\cxo {CXOU~J1647$-$4552}

\def\zerosei  {SGR~1806$-$20}
\def\xte     {XTE~J1810$-$197}

\def\zerozero {SGR~1900$+$14}
\def\ee        {1E~2259$+$586}

\def\galc {SGR 1729$-$45}
 \def \xmm {\emph{XMM-Newton}}
  \def \cha {\emph{Chandra}}
\begin{document}

\sloppypar
\title{Magnetars: properties, origin and evolution}

\titlerunning{Magnetars: properties, origin and evolution}        

\author{Sandro Mereghetti        \and 
            Jos\'e~A.~Pons \and   
            Andrew Melatos  
}

\authorrunning{S. Mereghetti,  J.A.~Pons, A. Melatos} 

\institute{S.Mereghetti \at
        INAF IASF-Milano, v. Bassini 15, I-20133 Milano, Italy\\
      \email{sandro@iasf-milano.inaf.it}
      \and  
       J.A.~Pons \at
     Departament de F\'{\i}sica Aplicada, Universitat d'Alacant,
    Ap. Correus 99, E-03080 Alacant, Spain\\
      \email{jose.pons@ua.es}
      \and
      A.~Melatos \at
      School of Physics, University of Melbourne, Parkville, VIC 3010, Australia\\ 
      \email{amelatos@unimelb.edu.au}
}


\date{Received: date / Accepted: date}

\maketitle

\begin{abstract}
Magnetars are neutron stars in which  a strong magnetic field is the main energy source. About two dozens of magnetars, plus several candidates, are currently known in our Galaxy and in the Magellanic Clouds. They appear as highly variable X-ray sources and, in some cases, also as radio and/or optical pulsars. Their spin periods (2--12 s) and spin-down rates ($\sim10^{-13}-10^{-10}$ s s$^{-1}$) indicate external dipole fields of $\sim10^{13-15}$ G,  and there is evidence that  even stronger magnetic fields are present inside the star and in non-dipolar magnetospheric components.  Here we review the observed properties of the persistent emission from magnetars, discuss the main models proposed to explain the origin of their magnetic field and present  recent developments in the study of their evolution and connection with other classes of neutron stars.
\end{abstract}

\section{Introduction}
\label{intro}

Magnetars are   neutron stars in which the main source of energy is provided by a strong magnetic field,  instead of rotation, accretion, nuclear reactions, or cooling.
While the bulk of rotation-powered (radio) pulsars have fields in the range B$\sim10^{11}-10^{13}$ G,  the external magnetic field of magnetars is typically   $10^{13}-10^{15}$ G and it is likely that their internal field is even stronger.
However, the distributions of field intensities for  magnetars and ``normal'' neutron stars overlap: there is not a discriminating B threshold between these two classes. Indeed, the presence of a strong dipole field (typically estimated from the star spin period and spin-down rate) is not a sufficient (nor a necessary) condition to trigger  ``magnetar-like'' activity. 
The latter is in fact mainly related to the presence of a significant toroidal component of the internal field, able to produce magnetospheric twists. 

Magnetars are   the most variable sources among the different classes of isolated neutron stars: their characterizing property is the emission, in  the X-ray and soft $\gamma$-ray range, of powerful  short  bursts  which often reach super-Eddington luminosities. More rarely, they also emit intermediate and giant flares, the latter involving the release of up to about 10$^{46}$  erg in less than half a second.  Magnetars also show  pulsed X-ray emission with typical luminosity  of $\sim10^{35}$ erg s$^{-1}$ in persistent sources,  and ranging from $\sim10^{32}$ to $10^{36}$ erg s$^{-1}$ in transient  ones.  The pulsations, caused by the neutron star rotation, have  periods of a few seconds which are secularly increasing on timescales from one thousand to several million years  ($\pdot \sim10^{-13}-10^{-10}$ s s$^{-1}$).

Due to this variety of phenomena, most of the sources that are now believed to be magnetars were initially classified in different ways and only later recognized as members of the same class of astrophysical objects. 
Bursts from magnetars had been observed since  the end of the 1970s \citep{maz79b,maz79}. 
They were initially classified as a sub-class of $\gamma$-ray bursts, with the peculiarity of  a softer spectrum and of coming repeatedly from the same sky directions \citep{nor91}.  They were thus named  soft $\gamma$-ray repeaters (SGRs). A secure identification with astrophysical objects known at other wavelengths was unfeasible with the  large positional uncertainties available at that time, but their possible association with supernova remnants\footnote{Ironically, it is now known that the nebulae associated to two of the three first discovered SGRs are not supernova remnants.}  suggested a neutron star nature.
Other sources that are now believed to be magnetars were  discovered as persistent pulsars in the soft X-ray range ($<$10 keV) and thought to be X-ray binaries powered by accretion, as most of the bright X-ray sources known at that time.
It was later pointed out that their narrow period distribution, long term spin-down,  soft X-ray spectrum and faint optical counterparts were at variance with the properties of  pulsars in massive binaries  \citep{mer95}. This led to their denomination as anomalous X-ray  pulsars (AXPs). 

We now believe that SGRs and AXPs are a single class of objects. In fact,   when   the persistent X-ray counterparts of SGRs were identified \citep{mur94,rot94,hur99,woo99c}, it was found that they are pulsating sources very similar to the AXPs \citep{kou98,kou99,esp09a,kul03}, and  SGR-like bursts were detected from several sources originally classified as AXPs \citep{gav02,kas03,woo05}.    
About two dozens of AXPs/SGRs are currently known in our Galaxy (plus one in each of the Magellanic Clouds)\footnote{An updated list   is mantained at   \texttt{http://www.physics.mcgill.ca/$\sim$pulsar/magnetar/main.html} \citep{ola14}.}. Most of them
show X-ray pulsations and have been seen to emit bursts. For   extensive reviews of the AXPs/SGRs observations and of the main   models proposed to   explain them see \cite{woo06}, \cite{mer08}, \cite{mer11a}, \cite{tur13}.  
 
We believe that  the most successful  explanation of the AXPs/SGRs is provided by the magnetar model \citep{tho95,tho96}, according to which  they are neutron stars powered by a strong magnetic field.
Alternative intepretations based on  isolated neutron stars  accreting  from fall-back disks formed  after the supernova explosion \cite[see, e.g.,][]{alp01,tru10} require some additional  process, besides accretion,  in order to  explain the powerful bursts and flares observed from  these sources.

In fact, the  suggestion that SGRs are neutron stars powered by magnetic energy  was first proposed to interpret the exceptional properties\footnote{The association with the supernova remnant N49 in the Large Magellanic Cloud yielded the distance and energetics of this event.} of the giant flare emitted by \lmc\ on March 5, 1979 \citep{pac92,dun92}.   In the following years, the original magnetar model has been considerably developed and expanded and it provides now the best explanation for the rich diversity of AXPs and SGRs phenomenology  \cite[see, e.g.,][]{bel11}. One essential feature of the magnetar model is the presence of significant twists in the magnetosphere \citep{tho02,bel07}, resulting in a structure quite  different from  that of the simple dipolar geometry assumed for normal radio pulsars and in magnetospheric currents with a charge density much larger than the classical Goldreich-Julian value. Bursts and flares can be  explained by  sudden releases of energy in the star interior leading to  fractures in the crust \citep{tho95}, by field reconnection events in the magnetosphere analogous to those occurring in the Sun \citep{lyu06a}, or by pair plasma fireballs produced by discontinuities in the propagation of fast MHD waves in the magnetosphere \citep{hey05a}.   

In Section \ref{properties}  we describe the properties of the so called  ``persistent'' emission\footnote{In the lack of a better nomenclature, we use this adjective somehow improperly also for transient and variable sources,  just to distinguish this emission  from that  of the short bursts  and of the intermediate/giant flares.} of magnetars. In the two following sections we review the scenarios that have been advanced for the formation of magnetars (Section \ref{origin}) and for their evolution (Section \ref{evolution}).

\section{Properties of the persistent emission}
\label{properties}

The main manifestations of magnetars occur in the X-ray energy range. All  known confirmed magnetars show pulsations in the soft X-ray band ($<$10 keV) and many of them have also been detected in hard X-rays, up to $\sim$100-200 keV. About half of the known magnetars have repeatedly been observed at nearly constant  X-ray luminosities of $\sim10^{34}-10^{35}$ erg s$^{-1}$,  with only  moderate variability (a factor of a few) on long timescales \cite[see, e.g.,][]{mer11a}. Much larger variability is seen in the transient magnetars, which reach the luminosity level of the ``constant'' magnetars only during outbursts lasting weeks/months and spend  the remaining time at a much fainter,  quiescent level,  $\sim10^{32}$ erg s$^{-1}$ or less \cite[see][and references therein]{rea11}.

The outbursts of transient magnetars are often associated with the emission of short bursts or flares \cite[e.g.,][]{woo05,esp08,mer09,apt09,van10}.   Short bursts are also emitted by the ``constant'' magnetars, but  they have never been detected from a transient magnetar in the quiescent luminosity level.

\subsection{X-ray pulsations}

The presence of regular pulsations with secularly increasing period, caused by the slowing down of the neutron star rotational velocity,  is one of the distinctive properties of magnetars and provides a very useful diagnostic tool for  their study. The 23 currently known magnetars have spin periods in a very narrow range (2-12 s), while their period derivatives span five orders of magnitude. Most of them have $\pdot$ in the range 10$^{-12}-10^{-10}$ s s$^{-1}$,  but in recent years a few  ``low-$\pdot$ magnetars'' have been discovered,  with spin-down rates as small as  4$\times10^{-15}$ s s$^{-1}$,  well in the range of those of rotation-powered pulsars \citep{rea10,rea12,rea13b,an13,sch14,rea14}.  
The observed distribution of   magnetars in the  pulsar $P-\pdot$ diagram  gives information on their evolution and relation with other classes of neutron stars.  
The lack of observed magnetars with periods longer than 12 s indicates that  the spin-down mechanism becomes highly inefficient at large ages and/or that old magnetars become more difficult to detect, for example  because their X-ray luminosity decreases and they emit bursts less frequently. The most obvious explanation to account for these effects is magnetic field decay \citep{col00,dal12}, as discussed in detail in Section \ref{evolution}.

\begin{figure}[htb]
\includegraphics[width=\columnwidth]{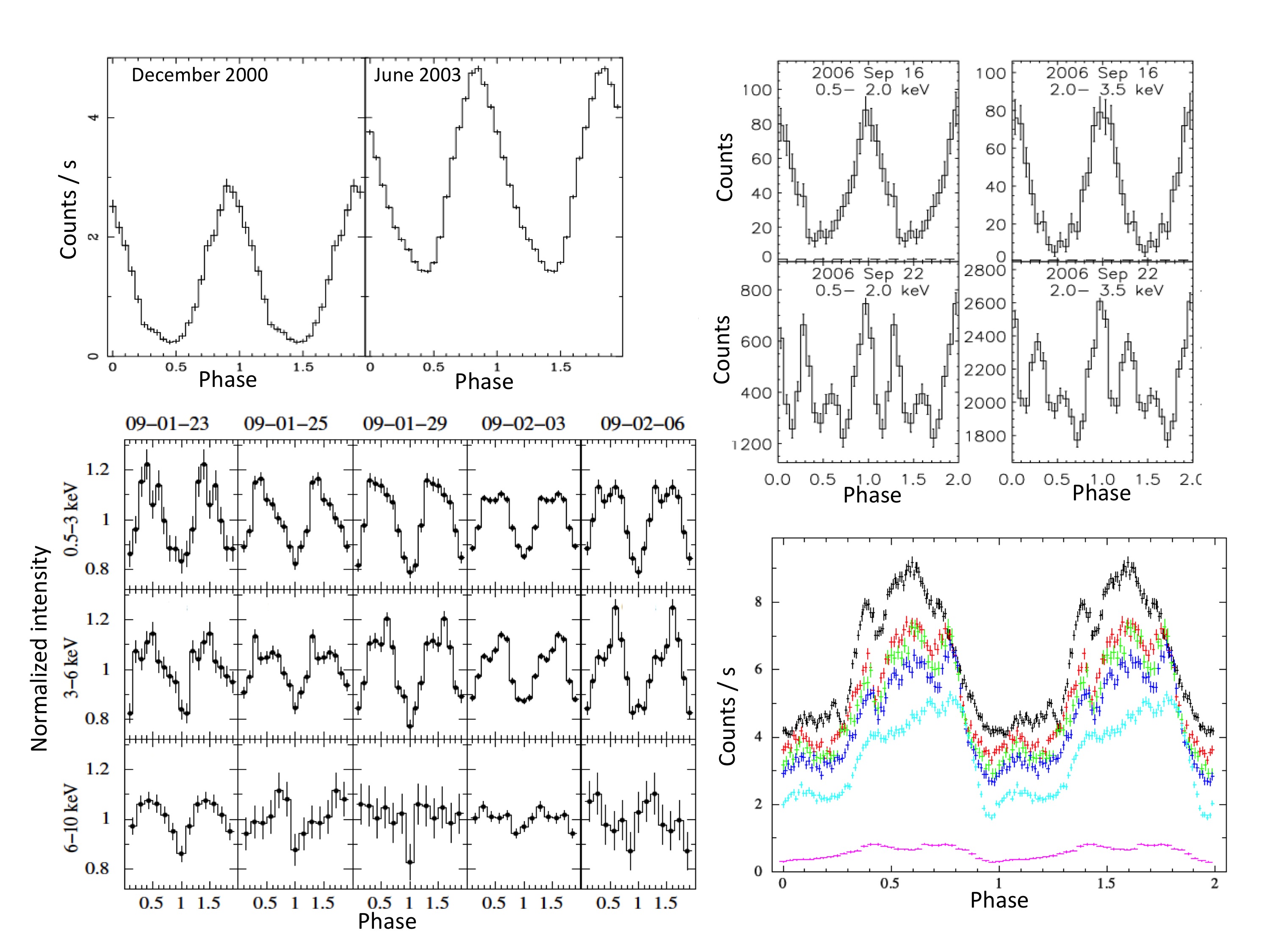}
\caption{Examples of X-ray pulse profiles of magnetars (for clarity, two neutron star rotations are shown in each plot). 
\textit{Top left panel:} the pulsed fraction of the persistent magnetar \oo\  anticorrelates with the luminosity \citep{mer04}:  when the flux was about twice that of the normal level, the pulsed fraction was smaller (53\% wrt 89\%);  both curves refer to the 0.6-10 keV range and were obtained with the EPIC instrument on \xmm .  
\textit{Top right panel:} pulse profile variations in two energy ranges of the transient magnetar \cxo\ \citep{mun07} in quiescence (\textit{upper panels}) and during the outburst (\textit{lower panels}).
\textit{Bottom left panel:} evolution of the pulse profiles of \qui\ during the decaying phase of the January 2009 outburst \citep{ber11}.
\textit{Bottom right panel:} pulse profiles of the transient magnetar \zeroc\ at different luminosity levels \citep{cam14}.
}
\label{fig-pulses1}
\end{figure}

\begin{figure}[htb]
\includegraphics[width=\columnwidth]{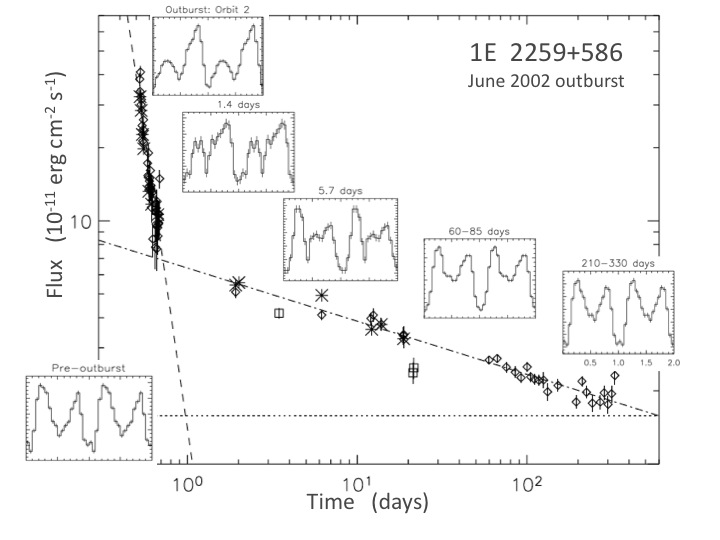}
\caption{Evolution of the pulse profile of \ee\ during the outburst of June 2002 (adapted from \citet{woo04}).  The insets show pulse profiles in the 2-10 keV obtained with the PCA instrument on $RossiXTE$. The inset in the lower left corner shows the pre-outburst pulse profile. Only the relative strength of the different peaks can be inferred from these profiles, which are plotted in arbitrary flux units.}
\label{fig-pulses2}
\end{figure}

Magnetars display  X-ray pulse profiles with a variety of shapes (from simple sinusoids to multipeaked) and spanning a large range of pulsed fractions (from less than 10\% to nearly 100\%). The pulse profiles are energy-dependent  (with a tendency toward more complex shapes with increasing energy) and, in many sources, time-variable. Changes in pulse profiles are  often connected with bursts/flares and/or glitches, but also long term variations, apparently unrelated to particular events, have been observed. Some examples of pulse profiles are shown in Figs.~\ref{fig-pulses1} and \ref{fig-pulses2}.

The spin-down of magnetars is attributed to the angular momentum carried away by  (time-variable) magnetized outflows and to dipole radiation losses. In a twisted magnetosphere, the latter effect produces a higher spin-down rate than in  pulsars with dipolar field because the twist inflates the poloidal lines and increases the magnetic field at the light cylinder.
Variations in the spin-down rate have been detected in practically all the magnetars for which good timing data extending over long periods are available. The $\pdot$ variations  are generally smaller than $\sim$50\%, but changes as large as a factor of ten over timescales of weeks have sometimes  been observed \citep{gav04b,dib09}. The variations in $\pdot$ observed in several magnetars are a proof of the dynamic nature of their magnetospheres. Since both the torque and the magnetospheric currents are  driven by  variations  in the twisting of the field lines \citep{bel09},  some correlations between the spin-down rate and X-ray emission properties can be expected, and indeed they have been observed (see Section \ref{correlations}).

Glitches have   been observed in many magnetars \citep{kas00,dal03,woo04,dib09}. They involve fractional frequency changes $\Delta\nu/\nu\sim10^{-7}-10^{-4}$, similar to those of the strongest glitches of radio pulsars \citep{dib08}, but the apparent paucity of magnetars glitches with smaller $\Delta\nu/\nu$ is  probably  a selection effect.  
The relation between magnetar glitches and  observed changes in the properties of their  X-ray radiation (e.g. bursts, flux or pulse profile changes, etc...) is extensively discussed in \citet{dib14}.  While the majority of radiative changes are associated with glitches (or with some other timing anomaly), the converse is not true:  several glitches seem to have no consequences on the emitted radiation.  
Two sources showed  episodes  in which the spin frequency apparently jumped in a short time interval to a value significantly smaller than that predicted by the observed spin-down rate. 
These events with negative  $\Delta\nu$  have been called antiglitches
and cannot be explained with the theories of normal glitches ($\Delta\nu>0$), which are based on the fact  that the neutron star interior rotates faster than the crust and acts as an angular momentum reservoir.
However, due to the sparse time sampling of the available  data, 
it is not clear if these events are really occurring on a short timescale, as normal glitches.
The first possible antiglitch, with $\Delta\nu/\nu=-10^{-4}$, coincided with the August 1998 giant flare of  \zerozero\   \citep{woo99b}.   
An analysis of the pulse phases suggests that the frequency variation could have been caused by an increase of more than a factor 100 in the spin-down rate in a few hours after the giant flare \citep{pal02}. 
However, an alternative possibility requiring only a factor of two  increase in the spin-down rate  in the three months preceding the giant flare cannot be ruled out.
A more convincing case  for a magnetar antiglitch,  
with  $\Delta\nu/\nu=-3\times10^{-7}$ and connected with a flux increase,  
has been recently reported   for \ee\ \citep{arc13}.
This prompted several interpretations, involving either a sudden event, like the collision with a solid body \citep{hua14},
or  a rapid reconfiguration of the magnetosphere with the opening of some of the closed field lines and the emission of an enhanced particle wind \citep{lyu14,ton14}.

\begin{figure}[htb]
\includegraphics[width=\columnwidth]{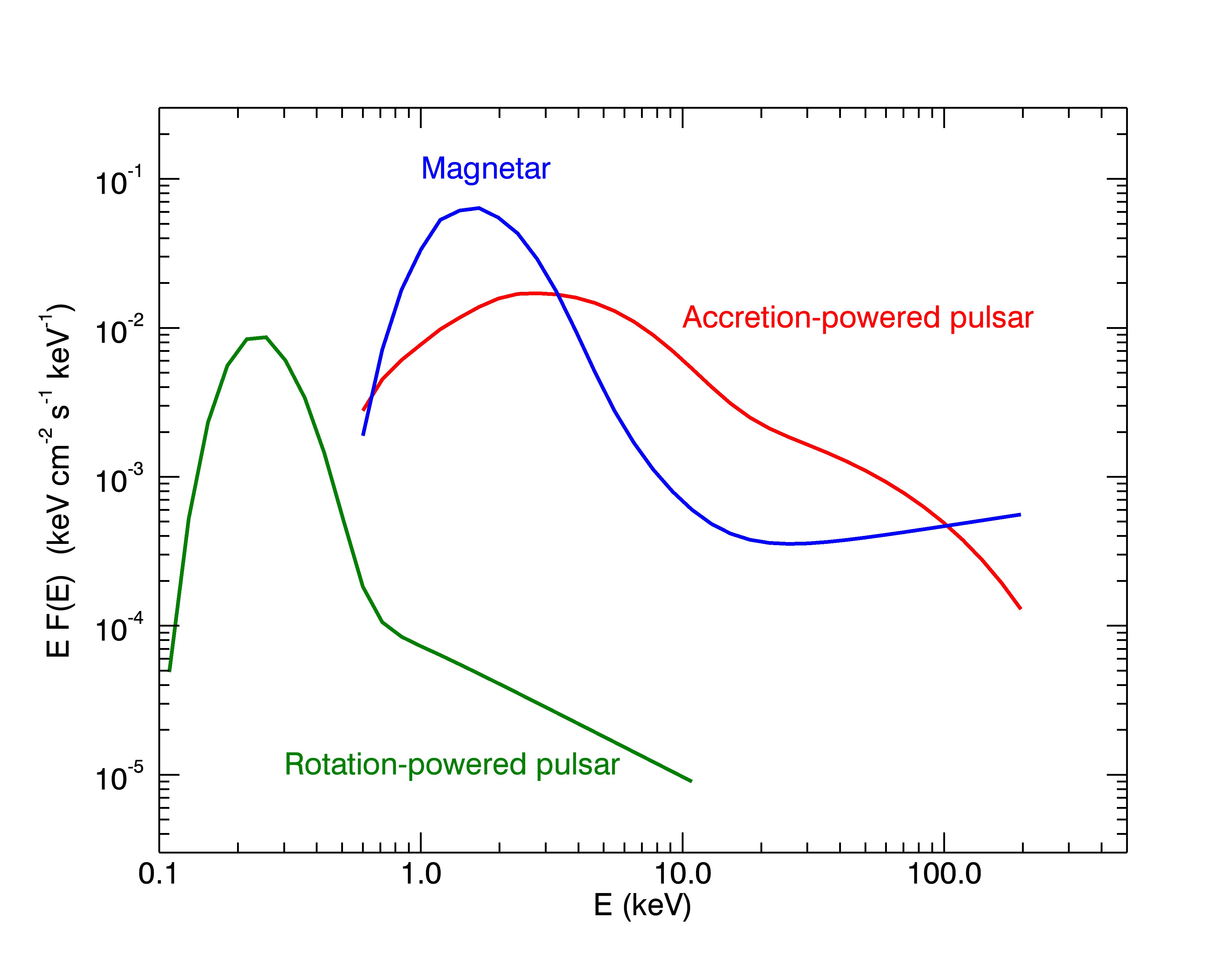}
\caption{Comparison of the X-ray   spectra of three  neutron stars representative of different classes. \textit{Red:} the accretion-powered binary X Persei (adapted from \citet{dis98}); \textit {Green:} the rotation-powered pulsar Geminga  (adapted from \citet{jac05}); \textit{Blue:} the magnetically-powered AXP \uu\  (adapted from \citet{rea07}).}
\label{fig-spectra}
\end{figure}

\subsection{X-ray spectra}
\label{xsp}

In Fig.\ref{fig-spectra} we compare representative spectra of three different classes of neutron stars: a magnetar (\uu ),  a rotation-powered pulsar (Geminga),  and an accretion-powered binary at low luminosity (X Persei).  If we limit the comparison to the $\sim$2-10 keV range, historically the first one to be  explored,  the most striking difference between these sources, is the softness of the magnetar spectrum. The first measured spectra of AXPs were in fact well fit by power-laws with  photon indexes $\Gamma$$\sim$3--4 \citep{par98,whi87,sug97}.  This characteristic spectral softness gives an immediate diagnostics to recognize magnetar candidates among newly discovered X-ray pulsars.  
When data of better quality became available, it was found  that a better phenomenological fit to the  magnetar spectra below 10 keV is provided by a blackbody model with temperature kT$_{BB}\sim$0.5 keV, plus either a power-law or a second blackbody component \citep{oos98,whi96,pat01,mer05c,mer06b}.
 
Contrary to the case of   rotation-powered neutron stars, for which the thermal and non-thermal components dominate in different energy ranges (see, e.g., Geminga in Fig. \ref{fig-spectra}), the higher temperature of magnetars implies that the blackbody and the power-law  contribute in a similar way to the 1-10 keV flux, making  more difficult to disentangle and constrain the two components.
The requirement of a power-law in the soft X-ray range might simply reflect the inadequacy of a simple blackbody to fit a more complex thermal model, rather than representing a physically distinct process. For this reason,  some caution is needed when drawing physical interpretations from some of the correlations between spectral parameters that have been reported in the literature.

Most magnetars are located at low Galactic latitude and thus their spectra are strongly affected by the interstellar absorption, with large column densities   N$_H\sim10^{22}-10^{23}$  cm$^{-2}$.  
The N$_H$ values required  by the blackbody plus power-law fits are often   
larger  than those independently estimated in other ways, suggesting that the power-law cannot be extrapolated to low energy without a cut-off. Good fits  are generally  obtained with the sum of  two-blackbody models, which 
can be interpreted in terms of  regions with different temperatures on the star surface \citep{hal05}.
Thanks to its location in the  Small Magellanic Cloud, \smc\ is the  magnetar with  the lowest interstellar absorption (N$_H\sim6\times10^{20}$ cm$^{-2}$) and offers the best opportunity to study  the X-ray emission at low energy:    its spectrum is well fit by the two-blackbody model while the power-law plus blackbody is rejected with high confidence  \citep{tie08}.

On the other hand, a power-law component is certainly present in the hard X-ray range.   
Several magnetars have been detected up to $\sim$150 keV with large pulsed fractions and spectra  flatter  than those of accreting X-ray pulsars. As schematically shown in Fig. \ref{fig-spectra}, the latter have exponential cut-offs  at a few tens of keV while the spectra of magnetars extend to higher energies.   
The first studies of the (non-bursting) emission from  AXPs/SGRs above $\sim$10 keV were   carried out  with the INTEGRAL,  RXTE and Suzaku satellites.  
Despite the limited sensitivity and  imaging capabilities of the instruments operating in this range, 
these observations were crucial to demonstrate that the hard X-ray  emission represents a non-negligible fraction of the energy output from magnetars \citep{kui04,mer05a,kui06,goe06b,den08a,den08b,eno10c,eno10}.
More sensitive observations  have been obtained in the last two years with the NuSTAR satellite, thanks to
the imaging capability provided by its focusing telescopes covering the 3--79 keV range. 
These observations allow to carry out spectral and variability analysis on short timescales  and to spatially resolve the hard X-ray emission in crowded and/or confused regions \citep{an13b,vog14,kas14}

We can summarize the  properties of the  hard X-ray emission from magnetars as follows:
\begin{itemize}

\item the luminosity in the hard component is similar to that observed below 10 keV.

\item  fits  in the range $\sim$10-200 keV with power-law models  give photon index values typically between   $\Gamma\sim1$ and 2 (except in the case of  \zerozero , which has $\Gamma\sim3$, \citet{goe06b}). 
    
\item  the flux upper limits derived in the MeV region \citep{kui06,den06} imply that the spectra cannot extend as power laws to such high energies.  Indeed, the data with high statistics show that curved models, like a log-parabolic function, provide better fits than simple power laws \citep{rea07,den08a,den08b}.  

\item   the spectra of the pulsed component are harder than those of the total emission  and show phase-dependent variations.

\item hard X-ray emission has been observed also in transient magnetars (\qui , \citet{eno10b}; \zeroc , \citet{rea09}; \galc , \citet{mor13,kas14}).   In the case of \qui\ the spectrum hardened as the flux decreased \citep{kui12}.

\item no detections at higher energy have been obtained\footnote{The MeV-GeV source in the region of \ee\ is well explained as emission from the supernova remnant CTB 109 interacting with molecular clouds \citep{cas12}.}. The upper limits derived with $Fermi$  in the  0.1-10 GeV range \citep{abd10,sas10} are incompatible with earlier predictions  which assumed emission from the outer magnetospheres of AXPs/SGRs \citep{che01,zha02}. 
Searches for TeV emission  with ground based telescopes gave negative results  \citep{ale13}.

\end{itemize}

The above description of the magnetar spectra is based on simple phenomenological fits, but in recent years   more physically-motivated models to interpret the observed broad band spectra have been developed \citep{tho05,bel09}.  Two main ingredients play an important role in these models: \textit (a) thermal emission from (a part of) the neutron star surface,  which is surrounded by a thin atmosphere, and \textit(b) the presence of a magnetosphere with a complex geometry and significant charge density. The magnetosphere affects significantly the emerging spectrum and  provides additional emission components due to the presence of accelerated charges.  
The surface thermal emission results from interior cooling powered by magnetic field dissipation and from external heating caused by  backward-flowing charges in the magnetosphere.
The presence of a relatively dense plasma in the magnetospheres with a twisted configuration, a distinguishing property of magnetars, has important implications for the emitted spectrum. Resonant cyclotron scattering of the thermal photons can produce hard tails \citep{tho02}.

First steps toward physical modeling taking into account the  effects of the
strong magnetic field on the thermal emission 
were  done considering radiation transfer in a mono-dimensional approximation  \citep{lyu06b,guv07}
More realistic 3-D computations required a Monte Carlo approach to study the photon propagation in a globally twisted magnetosphere supporting the strong currents that provide a large optical depth to  resonant cyclotron scattering \citep{fer07,nob08b}. These models  can successfully fit the observed spectra \citep{rea08,zan09}, but their validity above a few tens of keV is uncertain, because they are computed for non-relativistic (or only mildly relativistic) particle distributions
\footnote{The extension to the general case has been derived by \citet{nob08a},
but only limited applications have been reported \citep{zan11b}.}.
Also the assumed geometry  is probably oversimplified: the seed photons may have a non-uniform and time-variable temperature distribution over the star surface  and the magnetospheric twist may involve only a limited bundle of field lines rather than being global.   The modeling of phase-resolved spectra obtained during the evolution of transient magnetar outbursts is a promising approach which can give useful constrains on the models \cite[see, e.g.,][]{alb10}.
Unfortunately, phase-resolved spectra of good statistical quality are not always available, especially  in the hard X-ray range. 

A model for the hard X-ray spectra of magnetars has been developed by \citet{bel13} who numerically solved the radiative transfer of charged particles flowing in a large twisted magnetic loop. Relativistic electrons  ($\gamma$$\sim$10$^3$)  are injected  by high voltage discharges close to the star surface and flow in the closed loop (rather than along open field lines as in normal pulsars). The resulting emission, strongly beamed along the loop, has a hard power-law shape  below the MeV and is significantly suppressed at higher energy. This is  in agreement with the observations, but the model parameters depend strongly on the geometry of the loop and orientation of the source \citep{has14}. 
Both phase-averaged and phase-resolved spectra of magnetars recently obtained with the $NuSTAR$ satellite in the 4-80 keV range have been successfully fit by this model, which, however, does not yet include in a self-consistent way the low energy X-ray emission \citep{an13b,vog14}.

\begin{figure}[htb]
\includegraphics[width=\columnwidth]{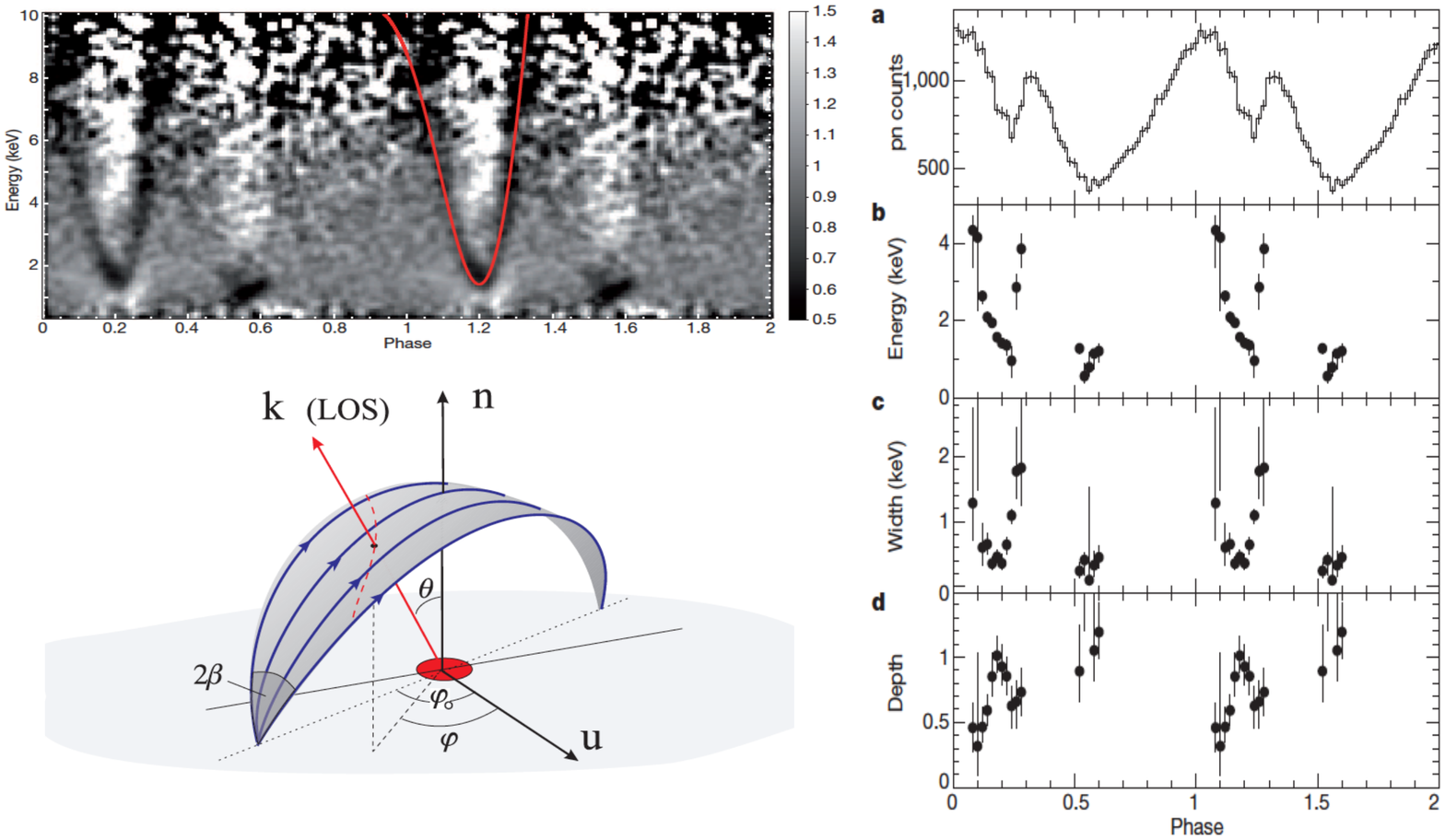}
\vspace{-2cm}
\caption{Phase-dependent line in the spectrum of the transient magnetar \zeroq\  discovered with an \xmm\ observation carried out on 2009  August 12, about two months after the beginning of the outburst (adapted from \citet{tie13}).  
\textit{Top left:} Phase-energy image obtained by binning the source counts into 100 phase bins and 100-eV-wide energy channels and normalising to the average spectrum and light curve. The red line indicates  (only on one of the two displayed cycles, for clarity) the expected phase-dependence of the line energy in the proton cyclotron  model illustrated in the next panel.   
\textit{Bottom left:} Schematic view of the model involving a magnetic loop over the X-ray emitting hot spot. The line of sight (LOS) intercepts the loop at different positions as the star rotates. The magnetic field varies along the loop, causing the observed shift in the line energy.  To reproduce the observed feature, the angle between  the rotation axis and the normal \textbf{n}  to the surface at the spot position  must be 20$^{\circ}$ and the LOS must form an angle of 70$^{\circ}$ with the rotation axis. 
\textit{Right:} Results of the phase-resolved spectroscopy. From top to bottom:  0.3-10 keV pulse profile folded at the spin period of 9.1 s (\textit{a}) ;  line energy (\textit{b}), width (\textit{c}), and depth (\textit{d}) of the cyclotron feature as a function of the spin phase.
}
\label{fig-0418}
\end{figure}

\subsection{Spectral features}

The detection of cyclotron lines is the most direct way to measure the magnetic field of neutron stars. This method has been successfully applied to accreting neutron stars in X-ray binaries since the beginning of X-ray astronomy \cite[see, e.g.,][]{rev14}. 
For magnetic fields of B$\sim10^{14}- 10^{15}$ G,  electron cyclotron lines are in the MeV energy range, where the currently available instruments are not sensitive enough to detect the magnetar emission. On the other hand, the energies of proton cyclotron lines fall in the soft  X-ray range, offering, in principle, a direct way to measure the magnetic fields of these objects. 
This motivated extensive searches for narrow features in the persistent emission from magnetars,
which however gave, until recently, only  negative results\footnote{Several features have been observed during bursts, e.g.: emission lines at $\sim$13-14 keV in \oo\ \citep{gav02,an14},  in \uu\ \citep{gav11}, and in \xte\ \citep{woo05}; an absorption line at 5 keV (and possibly its harmonics) in \zerosei\ \citep{ibr03}; an  emission line at 6.4 keV in \zerozero\ \citep{str00}.}. 
Some early claims with low statistical significance   \citep{iwa92,rea03} could not be confirmed with more sensitive observations \citep{rea05a}, implying   either line variability or spurious detection.  
The best upper limits, obtained with \xmm\ and \cha , yield equivalent widths smaller than a few tens of eV \citep{tie05a,jue02,tie08,rea09}.

Line smearing, caused, e.g.,  by the superposition of emission from regions of different field strength, is possibly one of the effects that reduce the detectability of cyclotron lines. Phase-resolved spectroscopy could mitigate this problem, but at the cost of a worse sensitivity due to the lower counts statistics of the spectra. 

The absorption line recently discovered in the transient magnetar \zeroq\  \citep{tie13} shows indeed a strong dependence on the star rotation phase and could be discovered only thanks to the examination of phase-energy images  (see Fig. \ref{fig-0418}). 
The line energy varies between $\sim$1 and $\gsim$5 keV within a small interval of the spin phase. This strong phase-dependence disfavours an explanation in terms of a  cyclotron line from electrons, which, given the dipolar field  B$_d$=$6\times10^{12}$ G inferred from the timing parameters $P$=9.1 and $\pdot$=4$\times10^{-15}$ s s$^{-1}$ \citep{rea13b}, should be at a height of a few stellar radii. 
A  field of 10$^{14}$ G at the star surface was  inferred by fitting the X-ray spectrum of \zeroq\ with a magnetic atmosphere model  \citep{guv11}.
As discussed in \citet{tie13},  the phase-dependent absorption line   is best interpreted as a cyclotron feature from protons residing  in a relatively small magnetic loop  with B$\sim(2-10)\times10^{14}$ G,  much higher than     B$_d$.
If this interpretation is correct, this  result, besides providing a direct estimate of the magnetic field strength close to the surface of a magnetar, confirms the complex topology of the magnetosphere, in which global and/or localized twists, as predicted by the magnetar model, play an important role.

\subsection{Radio emission}

Most magnetars  have not been detected in the radio band, despite being located above the death-line in the $P-\pdot$ diagram. The few magnetars detected in this band show radio properties very different from those of rotation-powered neutron stars.
Radio pulsations were first detected in two transient magnetars: \xte\ \citep{cam06} and \qui\ \citep{cam07c}. They are characterized by large  variability both in flux and pulse profile shape on timescale of days, by a very flat spectrum (S$_{\nu}$ $\propto$ $\nu^{\alpha}$, with $\alpha>-$0.5), and high polarization \citep{cam07a,cam08}. 
The first (and so far the only) magnetar discovered in the radio band,  \psr\ , was reported in 2010 \citep{lev10}. Its  radio properties are similar to those of the other radio-emitting magnetars. Although no clear signatures of magnetar-like activity have been seen in other wavelengths, its X-ray counterpart decreased in luminosity by  a factor over 50 from 2007 to 2011 \citep{and12}, suggesting that also in this case the radio emission is associated with a transient magnetar. 
The most recent addition to this small group is the transient magnetar \galc\ \citep{mor13}, which is particularly interesting due to its vicinity to the Galactic center. 
Its radio dispersion measure (DM=1770$\pm$3 pc cm$^{-2}$)   and Faraday rotation measure (RM=(--6.696$\pm$0.005)$\times10^4$ rad m$^{-2}$)  are  the highest among all known pulsars and indicate a distance very similar to that of the Galactic center  \citep{eat13,sha13}.  At this distance, the angular separation of 3$''$ between \galc\ and the Galactic center black hole SgrA*  corresponds to only $\sim$0.1  pc and thus  \galc\ has a non-negligible probability of being in a bound orbit with Sgr A* \citep{rea13}. By comparing the DM and RM values of  \galc\ with those of SgrA*,  and considering the density profile of the hot gas seen in X-rays,  \citet{eat13}  could constrain the magnetic field intensity at the beginning of the accretion flow onto the central black hole to be larger than $\sim$8 mG. If such a field is transported by the accretion flow it can be dynamically important for the accretion process on SgrA*.

The presence of radio emission gives the possibility to get very accurate positions and to measure proper motion through long baseline radio interferometry. The recently detected proper motion for   \galc\ suggests that this magnetar descends from one of the massive stars in the clockwise-rotating disk around the Galactic center \citep{bow15}. Proper motions have been measured  also  for \xte\  and \qui\ \citep{hel07,del12} in the radio band, while near IR observations yielded the proper motions of \zerosei\, \zerozero\, \ee\ and \uu\  \citep{ten12,ten13}).  These measurements correspond to transverse velocities  of $\sim$100-300 km s$^{-1}$,  not dissimilar from those of rotation powered  pulsars \citep{hob05}.

\subsection{Optical and infrared emission}

The study of magnetars in the optical and infrared is complicated by their intrinsic faintness at these wavelengths and by their location in strongly absorbed and crowded regions of the Galactic plane. Despite these difficulties, counterparts have been found for about one third of the known magnetars, and possible candidates have been suggested for a few other objects.  The associations are certain for the three sources showing optical pulsations:  \uu\ \citep{ker02,dhi05},  \oo\ \citep{dhi09}, and \zeroc\ \citep{dhi11}, while the other identifications are supported by the detection of long term variability  \citep{isr05,tes08,tam04}.

The   detected counterparts have magnitudes $\sim$23--26 in the optical band and  K$\sim$19-22 in the near infrared (NIR). They are    variable, but the relation between  the  optical and X-ray flux changes is unclear because only few truly simultaneous observations exist and different behaviors have been reported. Correlated variations were seen during the outbursts  of \xte\ \citep{rea04}, \ee\ \citep{tam04}, and \zeroc\ \citep{dhi11}, but also cases of apparently uncorrelated or anti-correlated variations were reported  \citep{tes08,cam07b,dur05c}.
The pulse  profiles of the three optically pulsed sources show a single broad peak, nearly aligned with the soft X-ray pulse,  and    pulsed fractions between $\sim$20\%  and  $\sim$50\%.

In the context of the magnetar scenario a few ideas for the origin of the optical/NIR emission have been put forward, involving non-thermal magnetospheric emission   \citep{eic02,bel07,zan11,bel13b}, but a detailed model is still lacking.

\subsection{Correlations}
\label{correlations}
 
\begin{figure}[ht!]
     \begin{center}
%
          \subfigure{%
            \label{cor_gamma1}
            \includegraphics[width=0.5\textwidth]{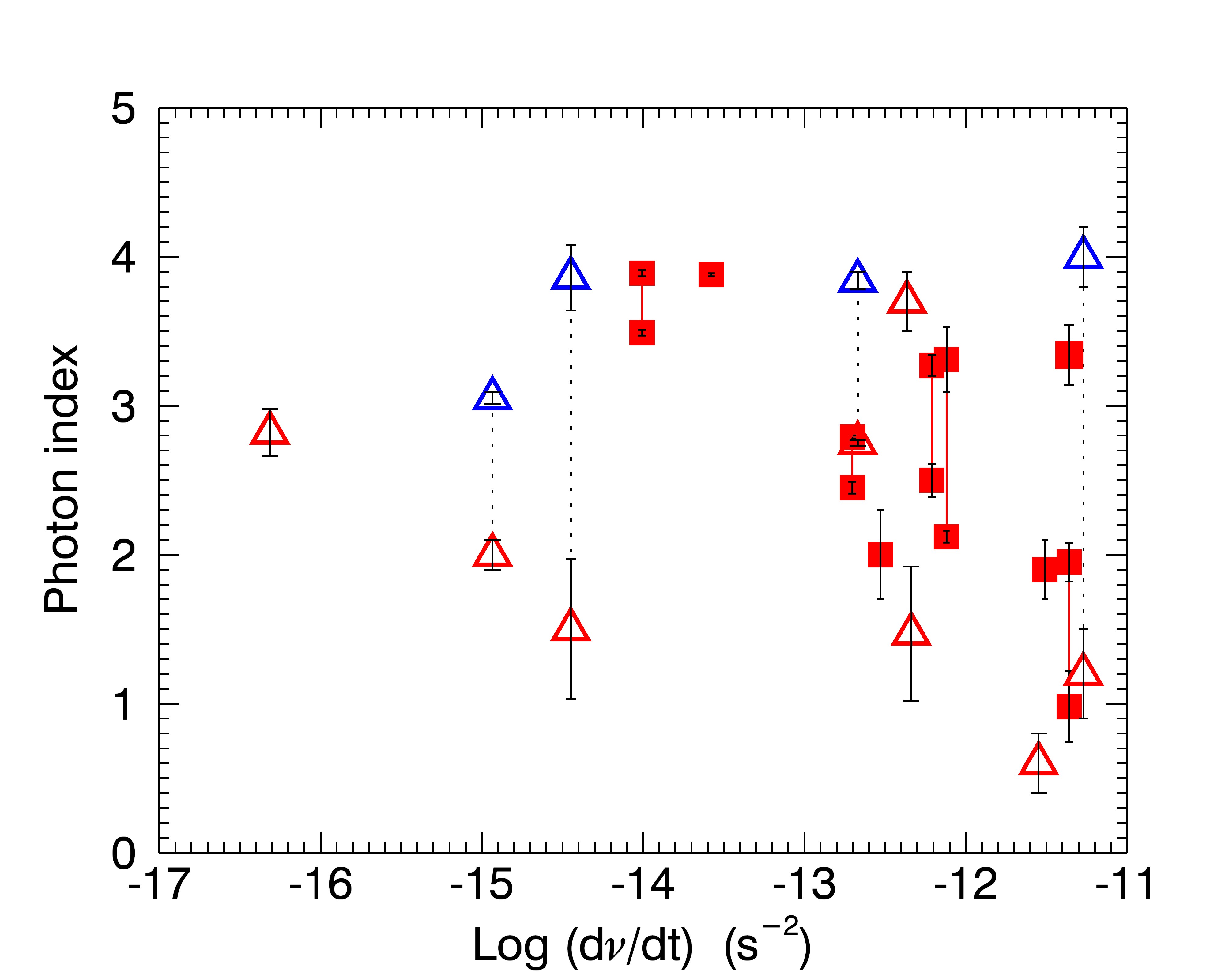}
        }%
        \subfigure{%
           \label{cor_gamma2}
           \includegraphics[width=0.5\textwidth]{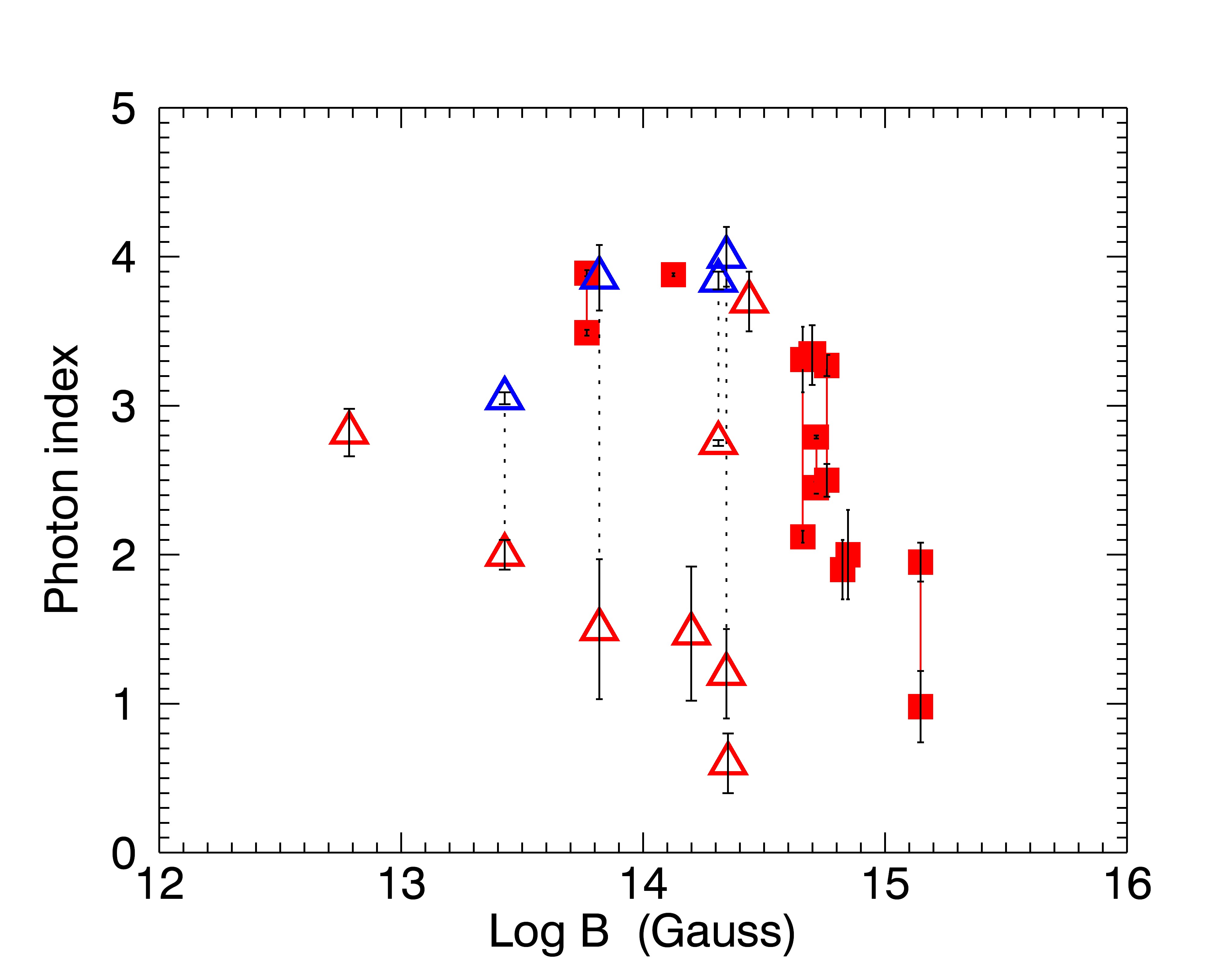}
        }\\ 
        \subfigure{%
            \label{cor_gamma3}
            \includegraphics[width=0.5\textwidth]{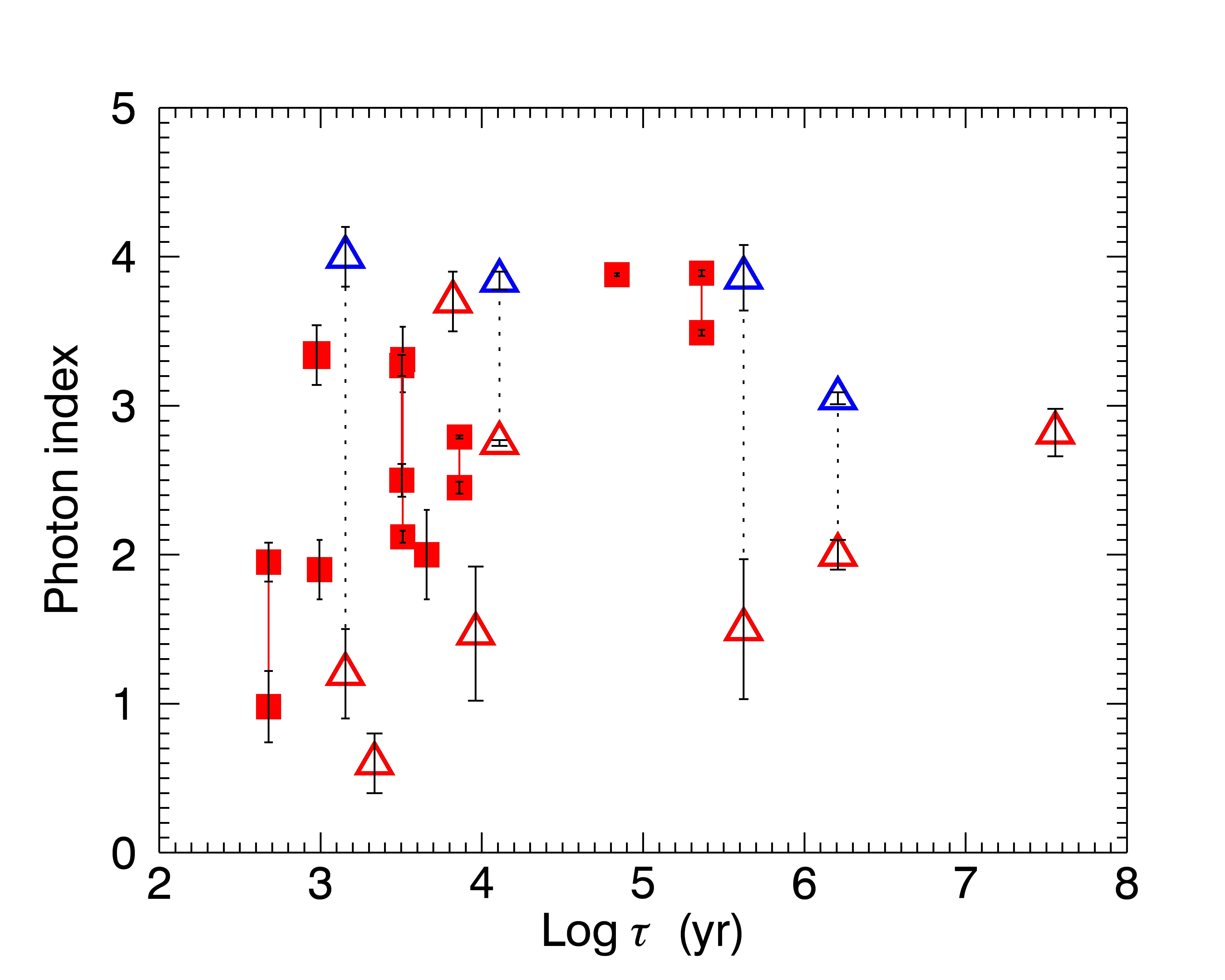}
        }%
        \subfigure{%
            \label{cor_gamma4}
            \includegraphics[width=0.5\textwidth]{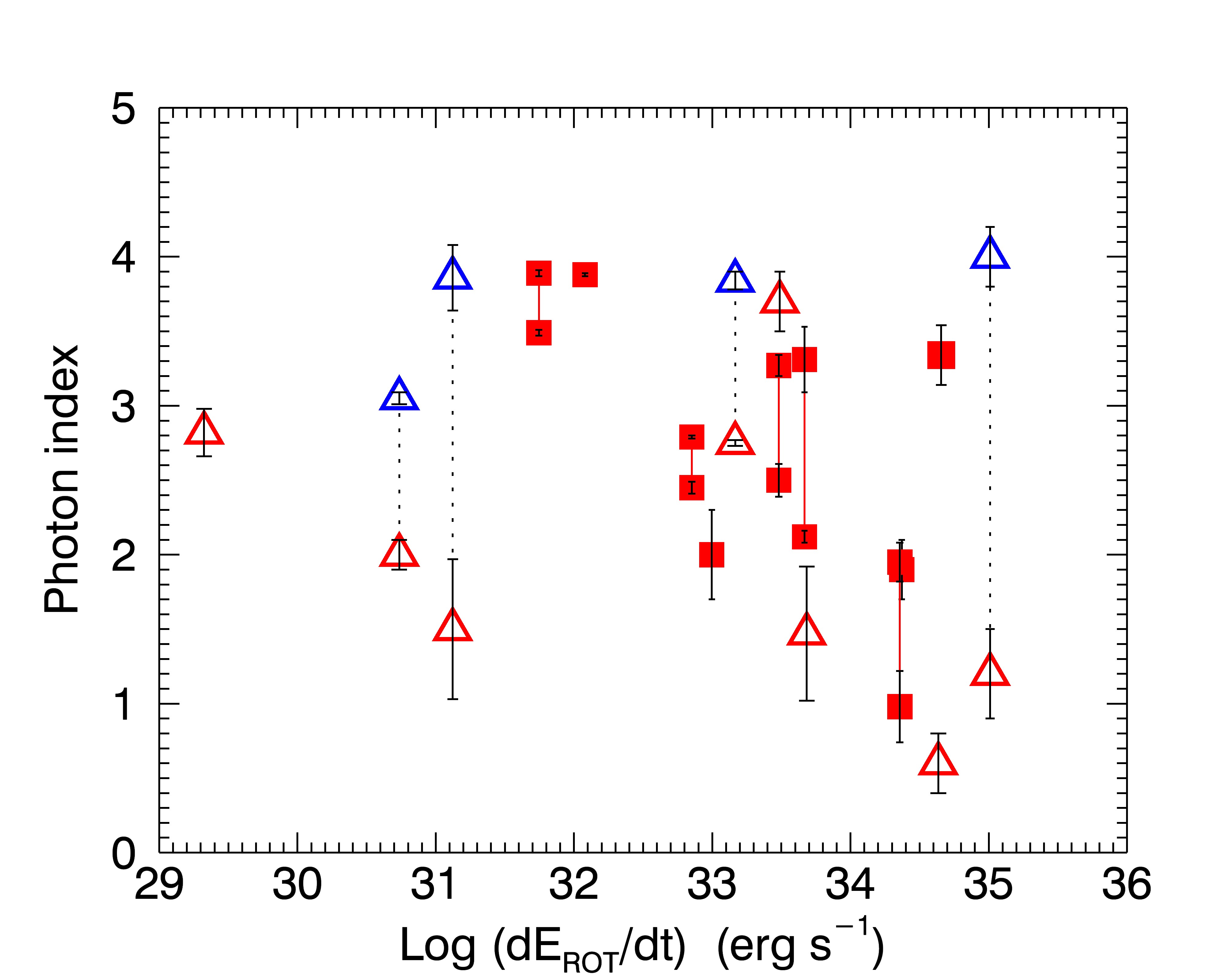}
        }%
    \end{center}
    \caption{%
        Dependence of the soft X-ray power-law photon index on different parameters:   $\nudot$ \textit{(top left panel)}, B$_d$  \textit{(top right panel)}, $\tau$ \textit{(bottom left panel)} and ${\dot E}_{rot}$ \textit{(bottom right panel)}.
Red squares indicate persistent sources (maximum and minimum observed values are reported for several sources). Triangles indicare transient sources during outburst \textit{(red)} and in during quiescence \textit{(blue)}.   }%
   \label{cor_gamma}
\end{figure}

Based on the small sample of seven  AXPs and SGRs  known at that time,  \citet{mar01}  pointed out that the sources with the larger spin-down rate have  smaller photon index in their soft X-ray spectra. The long term evolution of the power-law photon index and  $\pdot$ in \zerosei\ indicates that such a correlation between spectral hardness and average spin-down rate holds also for single sources \citep{mer05c}. An updated version of the photon index versus spin-down rate plot (top left panel of Fig. \ref{cor_gamma}) confirms the correlation for persistent sources (squares) and for transients in outburst (red triangles), but only for $\dot\nu \gsim 10^{-14}$ s$^{-2}$ and with some exceptions. The low-$\pdot$ sources have spectra harder than what would be expected from the correlation seen at higher spin-down rates, which also becomes less significant if the spectra of transients in quiescence are considered (blue triangles). In Fig. \ref{cor_gamma} we  show how the spectral hardness correlates with other quantities derived from the timing parameters, such as dipole field $B_d \propto (P\pdot)^{1/2}$, charactersitic age $\tau=P/2\pdot$, and spin-down power  ${\dot E}_{rot} \propto \pdot\ P^{-3}$.   As noted by \citet{kas10}, the best correlation is that with $B_d$. This is often considered to support the magnetar model: a stronger, and more twisted, field causes a larger spin-down rate as well as stronger magnetospheric currents which  harden the spectrum through resonant cyclotron scattering \citep{bar07}. However, the situation is probably more complicated, as also shown by the scatter of the points of Fig. \ref{cor_gamma},  and  other parameters might play an important role. As discussed in Section \ref{xsp}, it is difficult to disentangle the  thermal and non-thermal components in the soft X-ray range ($\lsim$10 keV) and  the particular geometry of the twisted bundles of magnetic field lines, not necessarily reflected in the derived  $B_d$ values, is the most relevant factor affecting the magnetar emission properties \citep{bel09}.

\begin{figure}[ht!]
     \begin{center}
%
          \subfigure{%
            \label{cor_kt1}
            \includegraphics[width=0.5\textwidth]{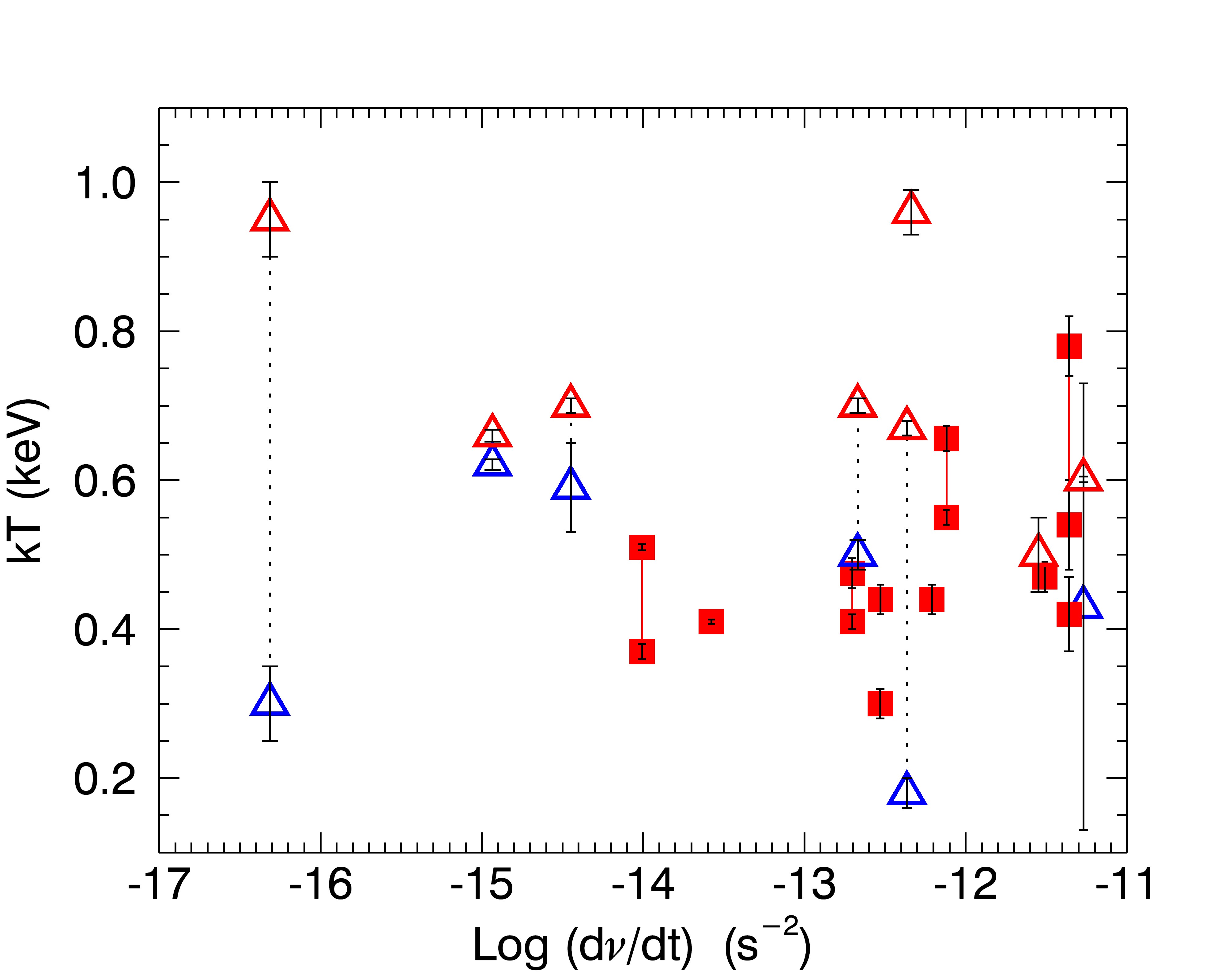}
        }%
        \subfigure{%
           \label{cor_kt2}
           \includegraphics[width=0.5\textwidth]{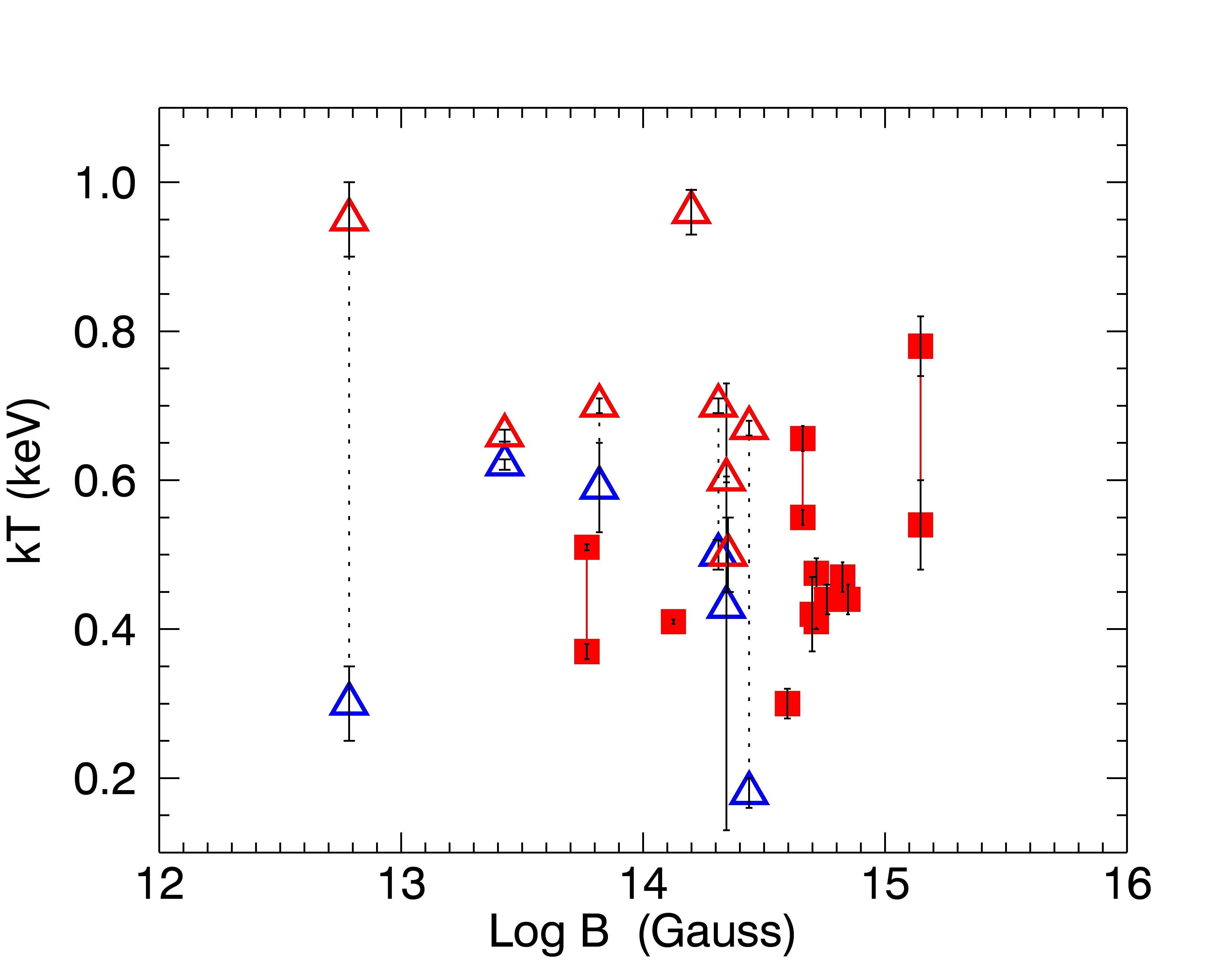}
        }\\ 
        \subfigure{%
            \label{cor_kt3}
            \includegraphics[width=0.5\textwidth]{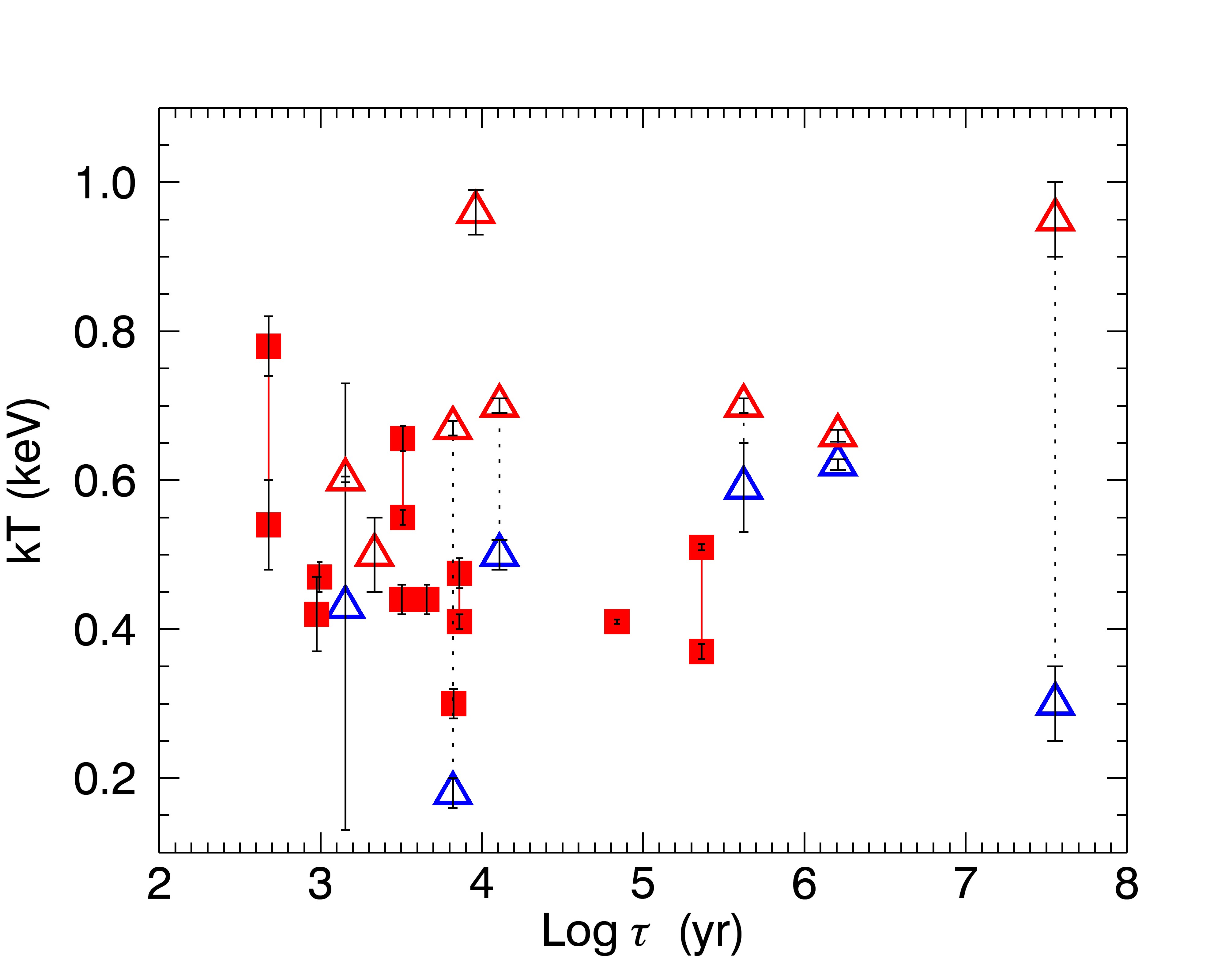}
        }%
        \subfigure{%
            \label{cor_kt4}
            \includegraphics[width=0.5\textwidth]{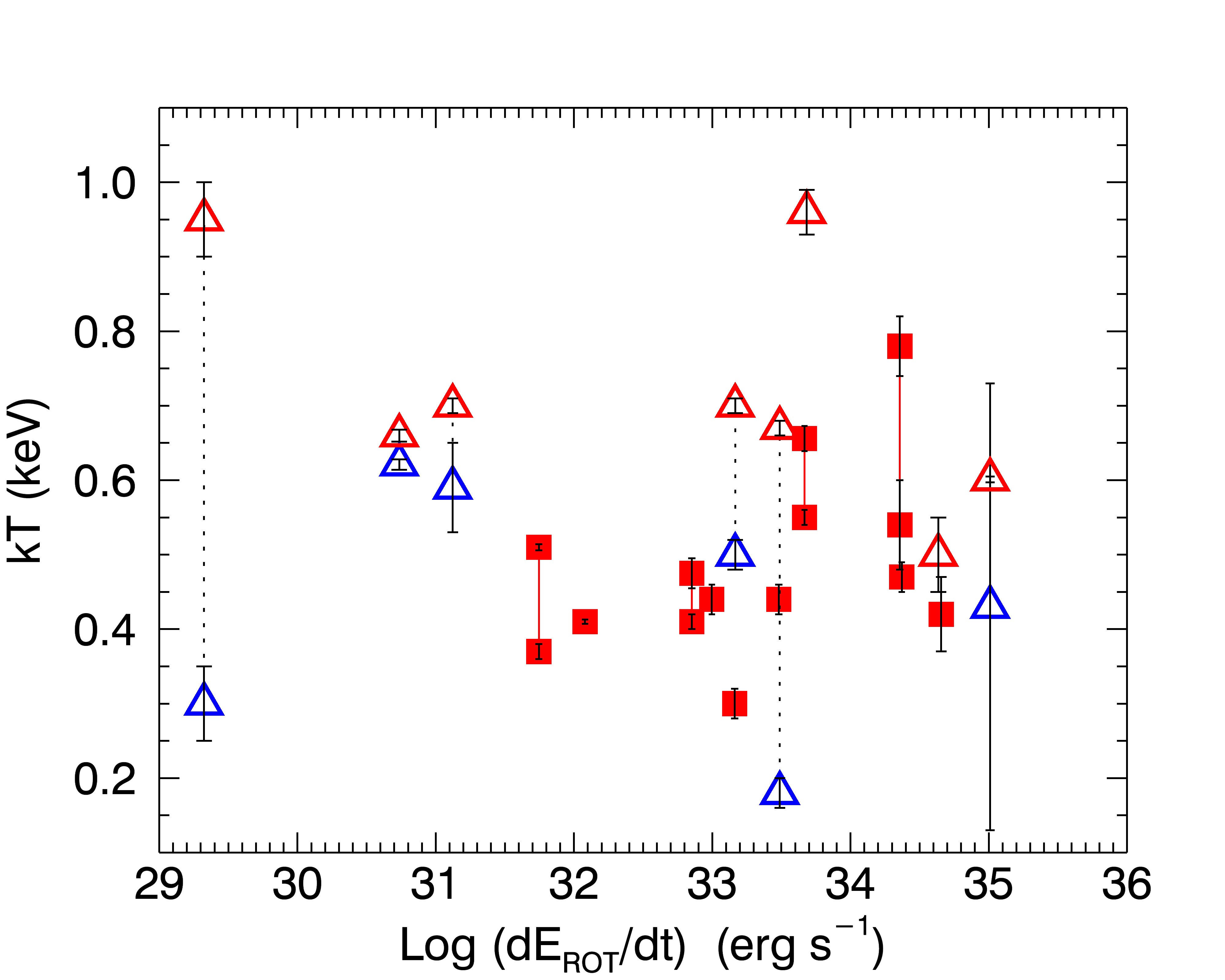}
        }%
    \end{center}
    \caption{%
                Dependence of the soft X-ray blackbody temperature on different parameters:   $\nudot$ \textit{(top left panel)}, B$_d$  \textit{(top right panel)}, $\tau$ \textit{(bottom left panel)} and ${\dot E}_{rot}$ \textit{(bottom right panel)}.
Red squares indicate persistent sources (maximum and minimum observed values are reported for several sources). Triangles indicare transient sources during outburst \textit{(red)} and in during quiescence \textit{(blue)}.   }%
   \label{cor_kt}
\end{figure}

Fig. \ref{cor_kt} indicates that the blackbody temperatures derived from double component fits to the soft X-ray spectra of magnetars do not show significant correlations with $\pdot$ or $B_d$ \citep{ola14}. However, when one compares the average temperature (or thermal luminosity) of magnetars with those of other classes of isolated neutron stars, interesting trends appear: there is a clear correlation between temperature and magnetic field \citep{aguilera08,ola13} and the magnetars are systematically more luminous than rotation-powered neutron stars of comparable characteristic age (see Section \ref{evolution}).
 
The first observations of persistent emission from magnetars above $\sim$20 keV revealed a  difference between the steep spectra of two SGRs with a remarkable record of bursting activity (\zerosei\ and \zerozero )  and those of the much quieter AXPs, which showed instead a significant hardening at high energy \citep{goe06b}. \citet{eno10c} reported an correlation between the hardness ratio (between the 20-100 keV and 2-10 keV fluxes) and $B_d$, based on the five persistent magnetars considered by \citet{goe06b} plus the transient \zeroc .   Similar conclusions were obtained by \citet{kas10}, who examined how the spectral turnover (defined as the difference between the  photon indexes of soft and hard X-rays)  correlates with spin-down rate and $B_d$.

\section{Magnetar formation and origin of magnetic field}
\label{origin}

Scenarios of magnetar formation need to reconcile two somewhat contradictory trends. On the one hand, the strong fields of magnetars argue for special conditions at birth: a highly magnetized progenitor in the fossil field hypothesis \citep{wol64}, or a rapidly rotating protoneutron star in the shear-driven dynamo hypothesis \citep{td93}. On the other hand, the magnetar birth rate ($\gtrsim$ 0.3 century$^{-1}$; \cite{kk08}) is comparable to rate of the core collapse supernovae  (1.9 $\pm$ 1.1 century$^{-1}$), measured by taking a $\gamma$-ray census of radioactive $^{26}$Al in the Galaxy\footnote{
The estimate from $^{26}$Al is subject to systematic uncertainties (included in the quoted error), as the isotopic yield is  model dependent, and there is an unknown yield contribution from local spallation processes and specific star-forming regions.
This rate is broadly consistent with extragalactic estimates \citep{dah12,tay14} which are undifferentiated by galactic type, e.g.\ comparative studies in the Local Volume which are limited  statistically by the small number of events \citep{bot12}.}
\citep{detal2006}. 
In this respect, the birth of a magnetar is not an unusual event, especially if there is  field decay (see Section \ref{evolution} and \cite{gh07}).

Current estimates of the natal magnetic field distribution of neutron stars in general, based on population synthesis studies \citep{fk06, kbm08}, Zeeman spectropolarimetry of progenitors \citep{letal07, wetal14}, and gravitational wave upper limits \citep{dss09, metal11, mm12}, are not sufficiently precise to pin down the shape of the distribution and test whether magnetars populate a second, high-field peak. In this section we present a brief summary of the advantages and disadvantages of the fossil and dynamo hypothesis for the origin of the magnetar magnetic field. The reader is referred to the paper by Ferrario et al. in this volume 
and to the  in-depth review by \cite{s09} for more details.

\subsection{Fossil field}

Magnetic flux conservation during the collapse of a massive progenitor, e.g. a chemically peculiar star with radius 3R$_{\odot}$ and a dipole magnetic field strength B$_0$ $\sim$ 10 kG, produces a natal neutron star magnetic field nominally as large as B $\sim 5\times10^{15}$G, enough to account for all known magnetars \citep{wol64, r72}. 
The fossil field scenario is therefore economical as it does not invoke a separate mechanism to produce magnetars, which  should  naturally derive  from the strong-field tail of the progenitor distribution. \cite{fw06} performed population synthesis calculations to show that the idea accounts also for the magnetization of strong-field white dwarfs, which exhibit similar mass-flux ratios. Population synthesis computations also predict that magnetars originate preferentially from the most massive O stars, consistent with some magnetars being associated with massive star clusters \citep{mun06,cla14,eik04,vrb00}
and with their very small scale height on the Galactic plane \citep{ola14}. Therefore magnetars should be  more massive than ordinary neutron stars, a claim which cannot be tested observationally at present.

The nominal maximum field B $\sim$ $5\times10^{15}$G implied by magnetic flux conservation is hard to attain for three reasons. First, only the central $\sim$ 2\% by cross-sectional area of the progenitor collapses to form a protoneutron star, reducing B proportionally. Second, there are too few progenitors with B$_0\gtrsim$ 10 kG to account for the magnetar birth rate inferred observationally \citep{kk08, woo08}. Third, magnetic core-envelope coupling in the progenitor brakes the core too efficiently to explain the observed neutron star spin distribution \citep{sp98} and leaves no room for the magnetic-dipole braking evolution normally envisaged for magnetars unless supernova kicks play a role. (Core-envelope coupling is an issue for the dynamo scenario too.)

\subsection{Protoneutron star dynamo}
Following flux compression, the magnetic field in a protoneutron star can be amplified further (over $\sim$ 10 s) by dynamo action driven by convection \citep{td93, bub05} or differential rotation \citep{b06, mbka06}. The relatively high ($\gtrsim$ 10\%) incidence of progenitors with B$_0\gtrsim$ 0.25 kG  from spectral class F0 to O4 \citep{wetal14} ensures that a seed field is available.

Neutrino-driven convection leads to protoneutron star fields of about 10$^{11}$G for a neutrino luminosity of $\sim 10^{44}$erg s$^{-1}$, which increases $\sim$10$^4$-fold by flux conservation when the protoneutron star collapses to form a magnetar \citep{td93}. However, these maximum field values are hard to attain, because dynamos typically operate at $\lesssim$ 5\% of equipartition \citep{css03, b06} and stratification quenches convection \citep{s09}.

In contrast, counter-intuitively, a shear-driven dynamo operates more efficiently under stratification \citep{b06}. The poloidal and toroidal field components grow in concert through the action of Tayler and/or magnetorotational instabilities \citep{b06, mbka06}, as well as r-mode instabilities \citep{cy14}; the same instabilities  also prevent premature saturation caused by back-reaction stress from the wound-up toroidal field. Differential rotation can also arise from binary mergers \citep{wtf14}, and mean-field magnetohydrodynamics (e.g. the $\alpha$ effect and anisotropic resistivity) and superfluid circulation to assist with amplification \citep{metal11, m12, gk13}. 

We do not expect to see evidence for rapid rotation at birth in the current magnetar population due to magnetic braking. However, a rapidly rotating protomagnetar is expected to power an energetic, relativistic wind for $\sim$ 10$^2$ s after birth, whose energy content is deposited in the supernova remnant. X-ray observations  of  three  supernova remnants associated with magnetars find no evidence for such ``over-powering", implying initial spin periods $\gtrsim$ 5ms \citep{vin06,mar14}, although this conclusion assumes an idealized Sedov expansion and neglects gravitational radiation which can nullify the over-powering issue \citep{dss09}.

In both dynamo scenarios, magnetic flux tends to escape buoyantly from the dynamo region \citep{rg92, s09, gk13}. Indeed, most of the flux would be lost via this process, were it not for helicity conservation, which stabilizes the situation for linked poloidal-toroidal fields under a variety of conditions \citep{bn06, lj12, aetal13, cr13,gou14}.

\section{Magneto-thermal evolution of magnetars}
\label{evolution}

In neutron stars endowed with strong magnetic fields, the temperature and
magnetic field evolution are closely inter-related. On one
hand, the dissipation rate of the magnetic field depends on the local value
of the electrical resistivity, which is a quantity
strongly dependent on temperature. On the other hand, the microphysics
ingredients determining the temperature evolution (heat capacity,
thermal conductivity, neutrino emission rates) are significantly modified by
the presence of a strong field.

A thorough and comprehensive discussion of all the  aspects involved in the magneto-thermal evolution of neutron stars can be found in the recent work by \citet{vigano13}, where the authors present results from 
simulations including two major novelties extending previous works \citep{GPZ99,PGZ00,pons07,aguilera08,pons09}: the proper treatment of the important Hall term in the induction equation describing the magnetic field evolution \citep{vigano12}, and updated microphysics inputs 
\citep[see][for  a review on matter properties in strong fields]{lai15}.
In this section, we briefly summarize the equations, the method, and the updated ingredients of the simulations.

\subsection{Basic equations}

For our purposes, the small structural deformations induced by rotation and   magnetic field can be safely neglected. To include general relativistic effects, we consider the standard static metric
\begin{equation}\label{eq:metric}
 ds^2 = - c^2 e^{2\nu(r)}dt^2 + e^{2\lambda(r)}dr^2 + r^2 d\Omega^2 ,
\end{equation}
where $e^{2\lambda(r)} = 1 - 2Gm(r)/c^2r$, $m(r)$ is the enclosed mass within radius $r$, and $\nu(r)$ is the metric factor accounting for redshift corrections. 

The neutron star magneto-thermal evolution is described by the coupled system formed by the
energy balance  and the Hall induction equations. The first reads:
\begin{equation}
\label{eq:heat_balance}
c_v e^\nu\frac{\partial T}{\partial t} - \vec{\nabla}\cdot[e^\nu \hat{\kappa}\cdot\vec{\nabla}(e^\nu T)] = e^{2\nu}(-{\cal Q}_\nu + {\cal Q}_h)
\end{equation}
where $c_v$ is the volumetric heat capacity, $\hat{\kappa}$ is the thermal conductivity tensor, ${\cal Q}_\nu$ are the energy losses by neutrino emission per unit volume, and ${\cal Q}_h$ is the Joule heating rate per unit volume. 
This is the first important coupling between the two evolution equations, because ${\cal Q}_h = \vec{j}^2 / \sigma$, where $j$ is the electrical current  determined by the magnetic field geometry and  $\sigma$ the electrical conductivity.
The second important effect of the presence of a strong magnetic field in the conduction of heat is the anisotropic conductivity tensor  ($\hat{\kappa}$), and the last one is that the magnetic field also affects the rate of neutrino processes, ${\cal Q}_\nu$. 

In the crust, ions form a Coulomb lattice, while electrons are relativistic, degenerate and can almost freely flow, providing the currents that sustain the magnetic field. The evolution of the magnetic field is governed by the Hall induction equation which, using the same notation as in \cite{pons09}, has the form:
\begin{equation}
\frac{\partial \vec{B}}{\partial t} = - \vec{\nabla}\times \left[ \frac{c^2}{4\pi \sigma} \curlB + \frac{c }{4\pi e n_e} \left[ \curlB \right] 
\times \vec{B}   \right]~
\end{equation}
where the conductivity $\sigma$ takes into account all the electron processes, which are strongly temperature-dependent, thus resulting in the strong coupling of the magnetic field evolution to the local evolution of temperature. 
The first term on the right hand side accounts for Ohmic dissipation, while the second term is the Hall term. 
The previous equation can be cast as
\begin{equation}\label{eq:induction}
\frac{\partial \vec{B}}{\partial t} = - \vec{\nabla}\times \left\{ \eta \left( \curlB + \omega_B\tau_e \left[ \curlB \right] \right)
\times \vec{B}   \right\}~
\end{equation}
where we have introduced the magnetic diffusivity $\eta = \frac{c^2}{4\pi \sigma} $ and the magnetization parameter $\omega_B\tau_e\equiv \frac{\sigma B}{c e n_e}$ (where $\omega_B=eB/m^*_ec$ is the gyration frequency of electrons, with $\tau_e$ and $m^*_e$ are the relaxation time and effective mass of electrons). This term regulates whether the evolution is dominated by the diffusive term or by the Hall term. In the regime where $\omega_B\tau_e\gg 1$ (strong magnetic fields, $\gtrsim 10^{14}$ G, and temperatures $\lesssim 5\times 10^8$ K, see \citealt{pons07,aguilera08,pons09,vigano12} for more details),  the Hall term dominates, and the induction equation acquires a hyperbolic character. The Ohmic and Hall timescales vary by orders of magnitude within the crust and during the evolution, depending strongly on density, temperature, and magnetic field intensity and curvature.
The main effect of the Hall term is to transfer part of the magnetic energy from large to small scales, as well as between poloidal and toroidal components.  In the case of strong toroidal components, it also leads to the formation of discontinuities of the tangential components of the magnetic field, i.e. current sheets, where the dissipation is strongly enhanced \citep{vigano12}. This directly affects the thermal evolution through the term ${\cal Q}_h$ in Eq.~(\ref{eq:heat_balance}).

The previous equations provide a proper description of the physics of the crust, once the microphysical input is provided. In the neutron star core, however, the situation is more complex.
The core of neutron stars (or at least a fraction of its volume) is thought to be a type II superconductor \citep{mig59,bay69}. The dynamics of the magnetic field in the core are not clearly understood. Standard Ohmic dissipation is irrelevant due to high conductivity, but other mechanisms such as the interplay between flux-tubes and vortices, magnetic buoyancy, or ambipolar diffusion may operate to expel magnetic flux from the core on timescales comparable to the thermal evolution timescale. The detailed study of these mechanisms 
is still lacking, and this explains  why most previous works   considered models with the field confined into the crust or used a very crude approach for the magnetic fields permeating the core \citep{hol02,pons07,pons09,gou14}.
Some basic issues as whether or not ambipolar diffusion plays any role at all are still under debate \citep{GJS11}.


\begin{figure}[htb]
\includegraphics[width=\columnwidth]{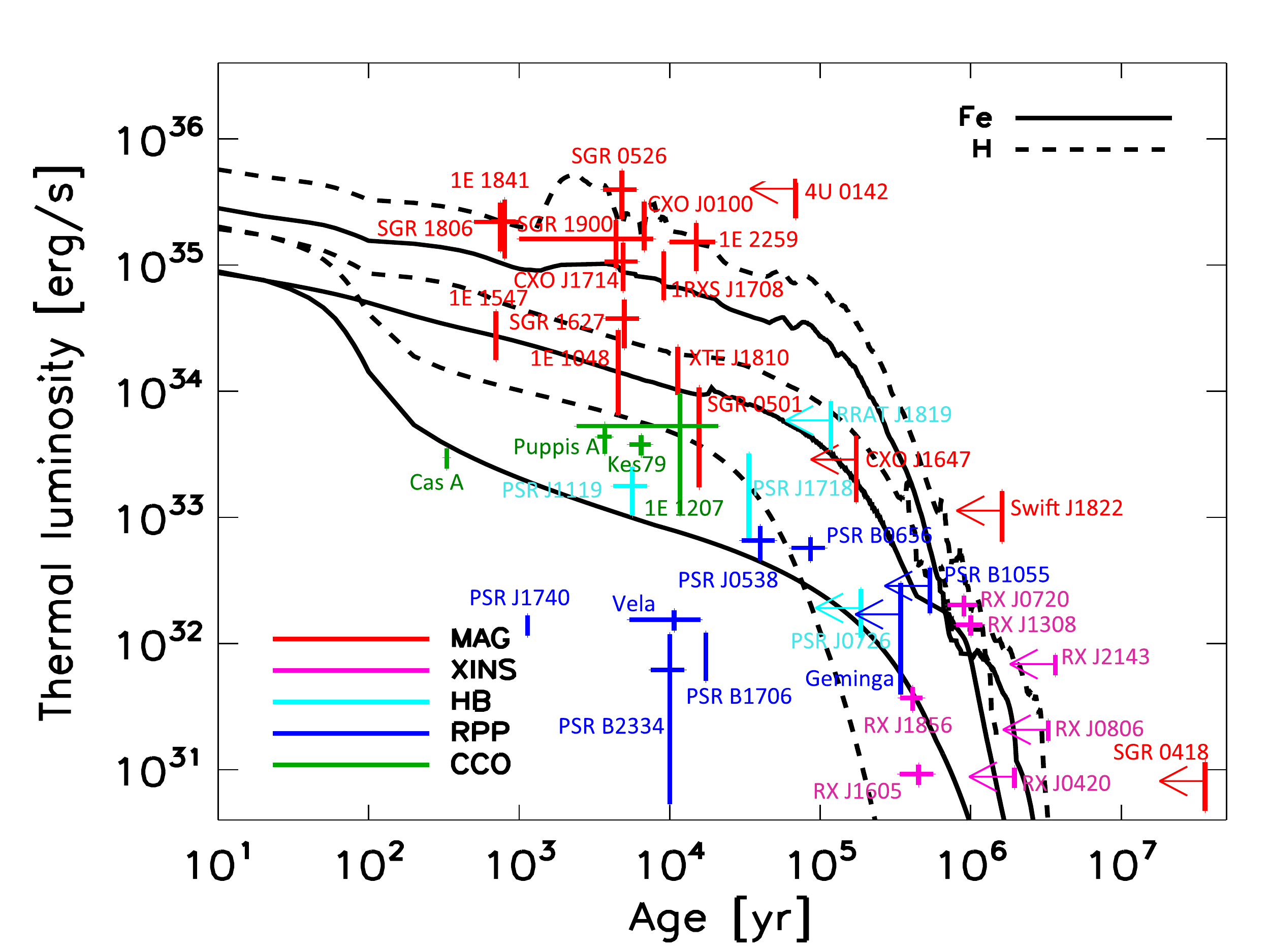}
\caption{Thermal X-ray luminosity versus characteristic age for magnetars (MAG) and other classes of isolated neutron stars (from \citet{vigano13}):  X-ray dim isolated neutron stars (XINS), rotation-powered pulsars with high (HB) and normal (RPP) magnetic field, central compact objects in supernova remnants (CCO) (see, e.g., \citet{mer11}, for a definition of these classes of sources and their X-ray properties). The lines indicate theoretical cooling curves for different compositions of the envelope and three values of the initial magnetic field in the crust ($3\times10^{15}$ G, $3\times10^{14}$ G  and 0 G, from top to bottom). A neutron star with  mass of 1.4 $\msun$ and radius 11.6 km has been assumed.}
\label{fig-cooling}
\end{figure}

\begin{figure}[htb]
\includegraphics[width=\columnwidth]{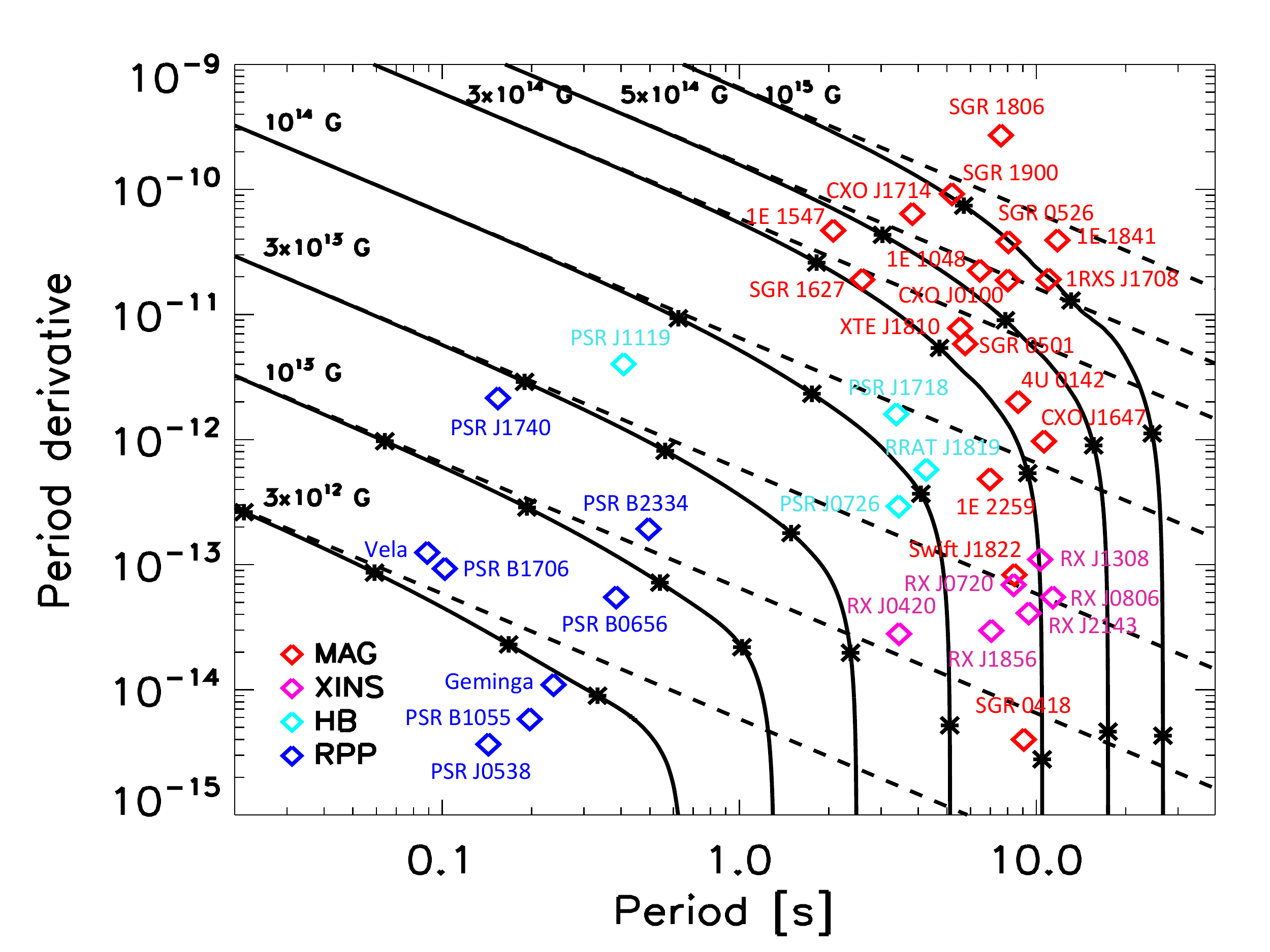}
\caption{$P-\pdot$ diagram for magnetars and other classes of isolated neutron stars (from \citet{vigano13}).  The solid lines indicate theoretical evolutionary tracks for different values of the initial magnetic field. The points corresponding to true ages of $10^3, 10^4, 10^5$ and $5\times10^5$ yrs are indicated by the asterisks on each lines. The dashed lines represent the evolution without magnetic field decay.}
\label{fig-ppdot}
\end{figure}

\subsection{Results}

From detailed numerical simulations solving the above system of equations, one can in principle obtain the local   temperature at each point of the neutron star surface and the integrated luminosity, to be compared to observations. But the reality is more complicated. Although both luminosities and temperatures can be obtained by spectral analysis,  it is usually difficult to determine them accurately. The luminosity is always subject to the uncertainty in the distance estimate, while the inferred effective temperature depends on the choice of the emission model (blackbody vs. atmosphere models, composition, condensed surface, etc.), and it carries large theoretical uncertainties in the case of strong magnetic fields.  It is often found that more than one model can fit equally well the data, without any clear, physically motivated preference for one of them.  Photoelectric absorption from interstellar medium further constitutes a source of error in temperature measurements, since the value of the hydrogen column density is covariant with the temperature value obtained in spectral fits. Different choices for the absorption model and the metal abundances can also yield different results for the temperature.
In addition, in the very common case of inhomogeneous surface temperature distributions, only an approximation with two or three regions at different temperatures is usually employed.
Moreover, in the case of data with few photons and/or strong absorption features, the temperature is poorly constrained by the fit, adding a large statistical error to the systematic one. 

With all the previous caveats in mind, the  studies of the magneto-thermal evolution of isolated NSs, have explored the influence of their  initial magnetic field strength and geometry, their mass, envelope composition, and relevant microphysical parameters such as the impurity content of the innermost part of the crust (the pasta region). The main findings can be summarised as follows (see Figures \ref{fig-cooling} and \ref{fig-ppdot}):

\begin{itemize}

\item
{\it Dependence on the magnetic field strength:}
The comparison between a range of theoretical models and the observations \citep{vigano13}, has shown that, for the objects   born with relatively low fields ($B_p\lesssim 10^{14}$~G), the magnetic field has little effect on the luminosity. Sources of this group,   which includes most of the ``normal'' radio pulsars,  have luminosities  which are compatible with the predictions of standard cooling models. The bulk of the magnetars, with  $B_p\sim$ a few $\times 10^{14}$~G  (as estimated from their timing parameters), display luminosities generally too high to be compatible with standard cooling alone. The magneto-thermal evolutionary models with $B_p^0\sim(3-5)\times10^{14}$~G can account for their range of luminosities at the corresponding inferred ages. As these objects evolve and their magnetic fields dissipate, their observational properties (both timing and luminosities) appear compatible with those of the XDINS\footnote{X-ray Dim Isolated Neutron Stars, a class of relatively old, purely thermally emitting neutron stars discovered by the $ROSAT$ satellite (see, e.g., \citet{tur09}, for a review of their properties).}. 
 
\item
{\it Relevance of the Hall term}: 
The Hall term plays a very important role in the overall magnetic field evolution, strongly enhancing the dissipation of 
energy over the first $\sim 10^6$~yr of neutron star life, with respect to the purely resistive case. This is due to two main effects: 
the generation of smaller structures and currents sheets, and the gradual compression of currents and toroidal field 
towards the crust/core interface. Hence, the rate of field dissipation strongly depends on the resistivity given by the
amount of impurities in the innermost region of the crust.  A highly impure or amorphous inner crust produces a significant increase in the field dissipation on timescales $\gtrsim 10^5$~yr.

\item
{\it Standard vs. fast neutrino cooling scenarios}: 
For weakly magnetized objects, low mass stars ($M \lesssim 1.4\,M_\odot$) are systematically brighter than high mass stars,  because fast neutrino processes are only
active above a certain threshold density. This separation is smeared out for highly magnetized stars,  for which the overwhelming
effect of the magnetic field makes it hard to distinguish if fast neutrino
cooling processes are acting or not, and therefore the mass dependence of
the cooling curves is much smaller.

\item
{\it Effect of the envelope composition:}
As a robust and general trend, light-element envelopes are able to maintain a higher luminosity (up to an order of magnitude) than iron envelopes for a long period of time, $\sim10^4$~yr, regardless of the magnetic field strength. 
The most luminous magnetars, with estimated field strengths $\sim10^{15}$~G from their timing parameters, are barely compatible with the $10^{15}$~G cooling curve with an iron envelope. However, for the same initial magnetic field, a light envelope is able to account for the luminosity of even the brightest objects.

\item
{\it Importance of the initial model}:
The initial magnetic field configuration plays a very important role in the observational properties of the NS. If the currents sustaining the magnetic field flow in the core,  their dissipation is negliglible, comparable with models in which (most of) currents flow in the crust. 
In particular, the presence of an initial strong dipolar, toroidal field in the crust breaks the symmetry with respect to the equator,
resulting in a hemisphere warmer than the other. If most of the initial currents are instead  confined into the core, then the reduced heat deposition in the crust results in a much cooler surface compared to the case in which the $B$ field lives in the crust only. 

\item{\it Influence in the magnetar outburst rate}:
The estimated outburst rate, resulting from breaking of the crust by the strong magnetic stresses, is found to be an increasing function of the initial magnetic field strength and a decreasing function of age (\citealt{perna11,pons11}).
A more quantitative comparison between the simulations and the observations is still not possible, due to the lack of sufficient statistics in the data. 

\end{itemize}

\section{Conclusions}

Although the currently known magnetars represent only a  small fraction of the observed neutron star population, they are attracting increasing interest, both from the observational and theoretical point of view. This is certainly due to their striking variability  properties, diversity of multiwavelength behavior and extreme physical conditions. In the last decades, they evolved from the status of poorly understood, exotic high-energy sources to become recognized as an important class of isolated neutron stars.
Although their  general properties  are well  explained in the context of the magnetar scenario, many aspects are still poorly understood and often the observational data are not sufficient to constrain the  model parameters. 
The transient nature of most of these sources implies that we have now discovered only a small fraction  of the magnetar population. The presence of wide field of view  instruments constantly monitoring the variable X/$\gamma$-ray sky is extremely important to further progress in this field exploiting the improved capabilities of future observational facilities.

\begin{acknowledgements}

We thank all the staff of the International Space Science  Institute and the organizers of the stimulating  Workshop ``The Strongest Magnetic Fields in the Universe''.   
The work of SM has been partially supported through the agreement ASI-INAF I/037/12/0.
JAP acknowledges support of the Spanish national grant AYA 2013-42184-P and of the New Compstar COST action MP1304.
AM acknowledges support of an Australian Research Council Discovery Project grant and is grateful to Nicole Darman for assistance with typesetting.
\end{acknowledgements}



\end{document}